\documentclass[aps,prl,reprint,amsmath,amssymb,superscriptaddress,longbibliography,floatfix]{revtex4-2}
\usepackage[T1]{fontenc}
\usepackage{lmodern}
\usepackage{amsthm}
\usepackage{graphicx}
\usepackage{booktabs}
\usepackage{longtable}
\usepackage{tikz}
\usetikzlibrary{calc,arrows.meta}
\usepackage[colorlinks=true,linkcolor=blue,citecolor=blue,urlcolor=blue]{hyperref}

\newcommand{\pair}[2]{\langle #1,#2\rangle}

\begin{document}
\title{Not All Holographic Entropy Inequalities \\Are Free of Area Law Ultraviolet Divergences}

\author{Xin-Xiang Ju}
\email{juxinxiang21@mails.ucas.ac.cn}

\author{Bart{\l}omiej Czech}
\email{bartlomiej.czech@gmail.com}

\affiliation{
Chen-Ning Yang Institute for Advanced Study,
Tsinghua University, Beijing, China
}

\begin{abstract}
\noindent
In the AdS/CFT correspondence, linear constraints on subsystem entropies, which are necessary conditions for the validity of the Ryu-Takayanagi proposal, define the holographic entropy cone. We disprove a prior claim that all holographic constraints of this type are free of area law divergences (superbalanced). We analyze one explicit counterexample and explain how to easily find more; we report signatures of $\mathcal{O}(10^5)$ counterexamples. Our findings invalidate multiple prior claims and conjectures.
\end{abstract}

\maketitle

\setcounter{secnumdepth}{1}
\section{Introduction}
\noindent
In the half century since Bekenstein \cite{Bekenstein:1973ur} and Hawking \cite{Hawking:1974rv} explained that black holes carry entropy, there emerged a fascinating connection between gravitation and quantum information theory. The holographic principle \cite{tHooft:1993dmi,Susskind:1994vu} posits that quantum information packed with Planckian density on a surface surrounding a gravitational system suffices to fully reconstruct the said system. In the most explicit version of holographic duality---the anti-de Sitter / conformal field theory (AdS/CFT) correspondence \cite{Maldacena:1997re}---extremal surfaces in the AdS spacetime (bulk) have a direct information theoretic interpretation on the CFT (boundary) side of the duality. In the simplest context---a time-reversal symmetric bulk geometry and CFT state---the Ryu-Takayanagi (RT) proposal \cite{Ryu:2006bv,Ryu:2006ef} equates the von Neumann entropy of CFT region $A$ with the area of the bulk minimal surface homologous to $A$:
\begin{equation}
S_A = (4G_N)^{-1} \min_{{\rm Area} \,\sim A} ({\rm Area})
\label{eq:rt}
\end{equation}
Multiple derivations of Einstein's equations \cite{Jacobson:1995ab,Lashkari:2013koa,Faulkner:2013ica} begin by assuming that area and entropy enjoy a simple relationship such as, for example, equation~(\ref{eq:rt}).
%Multiple derivations have traced the origin of Einstein's equations to a local relationship between area and entropy \cite{Jacobson:1995ab,Lashkari:2013koa,Faulkner:2013ica}.

This paper is concerned with necessary conditions, under which von Neumann entropies of CFT subsystems may manifest themselves as minimal areas, in accordance with the RT proposal (\ref{eq:rt}). The simplest example of such a necessary condition is the monogamy of mutual information \cite{Hayden:2011ag}:
\begin{equation}
S_{AB} + S_{BC} + S_{AC} - S_A - S_B - S_C - S_{ABC} \geq 0
\label{eq:mmi}
\end{equation}
All terms in (\ref{eq:mmi}) are von Neumann (entanglement) entropies of CFT subregions; concatenation denotes union ($AB \equiv A \cup B$). Inequality~(\ref{eq:mmi}) is a necessary condition for (\ref{eq:rt}) to work: states whose entropies violate (\ref{eq:mmi}) exist, but their entropies cannot be simultaneously represented as minimal areas in any bulk geometry. Following the pioneering work of \cite{Bao:2015bfa}, 85,706 analogous linear inequalities have been catalogued \cite{HernandezCuenca:2019wgh,Hernandez-Cuenca:2023iqh,HernandezCuenca:HECDatabase},
as well as two infinite parametric families \cite{Czech:2023xed, Czech:2024rco}. All of these {\bf holographic entropy inequalities} are necessary conditions for identifying CFT subregion entropies with minimal surface areas according to RT. The values of subsystem entropies, which are achievable in quantum mechanics and allowed by the holographic entropy inequalities, are said to span the {\bf holographic entropy cone} \cite{Bao:2015bfa}.

We are concerned with the ultraviolet (UV) finiteness of these inequalities. The entanglement entropy of a CFT subregion $A$ is UV-divergent, with the leading divergence proportional to the area $\partial A$ \cite{Bombelli:1986rw,Srednicki:1993im}. On the AdS side of holographic duality, this is encapsulated by the infrared (IR) divergence of areas of minimal surfaces, which extend out to the asymptotic boundary of spacetime. A remarkable feature of the monogamy of mutual information is that although area law divergences are present in every term, they cancel out from inequality~(\ref{eq:mmi}). For this very reason, the left hand side of (\ref{eq:mmi})---known as minus `topological entropy'---is useful for classifying two-dimensional spin systems~\cite{Kitaev:2005dm,Levin:2006zz}. Even more remarkably, the same cancellation of area law divergences is true in the two infinite families and all the 85,706 other hitherto known holographic entropy inequalities. This suggests that they too might be useful for characterizing `topological entanglement' in two-dimensional systems~\cite{Bao:2015boa}. 

It has been claimed \cite{He:2020xuo} that the cancellation of area law divergences---known as {\bf superbalance}---must extend to all holographic entropy inequalities, including the not-yet-discovered ones. Heuristically, one expects that the UV divergences of CFT entanglement entropies are hard-wired into the asymptotic boundary conditions of the AdS spacetime, so the RT proposal should not be vulnerable to UV features of the CFT state. By now, the claim that all holographic entropy inequalities are superbalanced is deeply engrained in the literature. 
%and has become second nature to many researchers.

The present paper refutes this claim. One counterexample is closely related to inequality~(\ref{eq:witnessineq}). We report signatures of some $\mathcal{O}(10^5)$ other counterexamples, a number comparable to the total count of known holographic entropy cone features in database~\cite{HernandezCuenca:HECDatabase}. 
%For technical statements, see the text below. 
%total count of holographic entropy inequalities (outside the infinite families).
%In the rest of the paper, we explain the initial rationale for conjecturing that holographic entropy inequalities should be superbalanced, develop an argument that leads to a counterexample, and discuss the consequences. 
Our findings invalidate a sizeable portion of what we thought we knew about the holographic entropy cone.
\smallskip

{\it Note on nomenclature---} All subsystem entropies must obey the subadditivity inequality \cite{Araki:1970ba}, which states that the mutual information $I(A:B)$ is non-negative:
\begin{equation}
I(A:B) \equiv S_A + S_B - S_{AB} \geq 0
\label{eq:sa}
\end{equation}
Condition~(\ref{eq:sa}) does not tell us whether a quantum state is inconsistent with the RT proposal; it tells us that some putative states do not exist at all. For this reason, in this paper we do not consider subadditivity to be a \emph{holographic} inequality. To keep the distinction between holographic and non-holographic inequalities meaningful, we likewise do not consider the combination (\ref{eq:mmi}) + (\ref{eq:sa}) to be holographic. 

This choice of nomenclature is useful because inequality~(\ref{eq:sa}) is not superbalanced (see equation~(\ref{eq:bellsa}) below), and neither is the convex combination (\ref{eq:mmi}) + (\ref{eq:sa}). The claim of superbalance \cite{He:2020xuo}, which the present paper refutes, was intended to apply to holographic inequalities in the sense specified here---for inequalities with no admixture of subadditivity.

\section{A useful notation and facts\\ about entropy inequalities}
\noindent
An assignment of subsystem entropies can be represented by a vector $s$ with components $S_A$, $S_B$, \ldots, $S_{AB}$, $S_{AC}$\dots\, The coefficients of any entropy inequality such as (\ref{eq:mmi}) or (\ref{eq:sa}) can likewise be represented by a vector $e$. When we test whether an entropy vector $s$ respects an inequality $e$, we substitute its entropies into the inequality to arrive at a real number, which we denote $\langle s, e \rangle$. The space of entropy assignments and the space of inequality coefficients are dual to each other, in the sense that we have a natural bilinear pairing $\pair{\ldots}{\ldots}$. 

By definition, all holographic entropy inequalities obey the following condition:
\begin{itemize}
\item If the entropy vector $s$ is RT-realizable then:
\begin{equation}
\pair{s}{e} \geq 0
\label{eq:pairing}
\end{equation}
`RT-realizable' means that the component entropies of $s$ can be consistently realized as areas of minimal surfaces in a single bulk geometry. 
\end{itemize}
In addition, for every vector $s$ whose component entropies are consistent with quantum mechanics, we have:
\begin{equation}
\langle s, i_{AB} \rangle \geq 0 \qquad {\rm and} \qquad \langle s, i_{JK} \rangle \geq 0 
\label{eq:sa-rewrite}
\end{equation}
Here $i_{AB}$ is the vector of coefficients of inequality~(\ref{eq:sa}): $+1$ for the coefficients of $S_A$ and $S_B$ and $-1$ for the coefficient of $S_{AB}$. Notation $i_{JK}$ refers to all instances of subadditivity $I(J:K) \geq 0$, where regions $J$ and $K$ may be composite, e.g. $J = AB$.

Area law divergences of CFT subregion entropies, which are the focus of this paper, can be modelled in a simple way. If regions $A$ and $B$ share a common boundary in the CFT, its area makes a UV-divergent contribution to the entropy of any region that contains exactly one of $A$ and $B$, but does not affect entropies of regions that contain both $A$ and $B$ or neither. This characterization is identical to a Bell pair shared by $A$ and $B$~\cite{Bao:2015boa}. Indeed, in the original derivation~\cite{Srednicki:1993im}, the divergent area-law entanglement arises from Bell pairs that straddle subregion boundaries, integrated over all scales.
%The Bell pair origin of the area law is manifest in derivation~\cite{Srednicki:1993im}, in which the divergence arises from Bell pairs that straddle subregion boundaries, integrated over all scales.
%where removing the ultraviolet cutoff yields a divergent contribution composed of Bell pairs from all scales. 

In what follows we model area law divergences with Bell pair entropy vectors $b_{AB}$. This entropy vector has $+1$ is all terms containing exactly one of $A$ or $B$ and $0$ in all terms containing both or neither. We note the following facts:
\begin{itemize}
\item An inequality $e$ is superbalanced if for all atomic (not composite) regions $X,Y = O, A,B,C\ldots$
\begin{equation}
\langle b_{XY},\, e\rangle = 0
\end{equation}
This includes a potential purifying system $O$, which may be evoked without loss of generality so that the global state on $OABC\!\ldots\,$ can be assumed pure.
\item Subadditivity~(\ref{eq:sa}) is not superbalanced:
\begin{equation}
\label{eq:bellsa}
\langle b_{XY},\, i_{XY} \rangle = 2
\end{equation}
\end{itemize}
%Assuming Haag duality holds, complementary regions in pure states have equal entropies. For example: 
%\begin{equation}
%S_{C} = S_{DO}\quad \textrm{in a pure state on $OAB$}
%\end{equation}
%This allows us to rewrite every inequality in many ways. For example, subadditivity~(\ref{eq:sa}) on a pure state on $OAB$ can be rewritten as $S_{DO} + S_D - S_O \geq 0$. 
%%\begin{equation}
%%S_{DO} + S_D - S_O \geq 0
%%\label{eq:sa2}
%%\end{equation}
%So long as one atomic subregion $X$ is consistently designated as the purifier---that is, each term is written in the way that excludes $X$---then superbalance takes a simple form: 
%\begin{itemize}
%\item For every pair of atomic regions $X$ and $Y$, the net number of terms that contain $X$ alone, $Y$ alone, and $XY$ together must be zero. 
%\end{itemize}
%This follows directly from the properties of bell vectors $b_{XY}$. 
%Observe that---in agreement with the claim of \cite{He:2020xuo}---ineq.~(\ref{eq:mmi}) is superbalanced whereas subadditivity is not.

\section{Non-superbalanced\\ holographic inequalities}
\noindent
Our goal is to find a holographic entropy inequality, which is not superbalanced. That is, we seek an $e$ with:
\begin{equation}
 \pair{b_{AB}}{e}>0.
 \label{eq:criterion-bell}
\end{equation}
To confirm that $e$ is holographic, we must verify: 
\begin{enumerate}
\item Criterion~(\ref{eq:pairing}): $e$ is valid on all RT-realizable states. 
\item Inequality~$e$ cannot be written as a convex combination of a superbalanced holographic inequality $e'$ and instances of subadditivity.
\end{enumerate}

The first point is routine. It is commonly known \cite{Bao:2015bfa} that criterion~(\ref{eq:pairing}) follows if one can find a requisite {\bf contraction map} $f$. Writing $e$ in the form $\sum_{a=1}^L \alpha_a S_{J_a} \geq \sum_{b=1}^R S_{K_b}$, we need an $f: \{0,1\}^L \to \{0,1\}^R$, which satisfies the following two conditions:
\begin{itemize}
\item Contraction condition. For all $x, x' \in \{0,1\}^L$:
\begin{equation}
\sum_{a=1}^L \alpha_a ||x_a - x'_a || \geq \sum_{b=1}^R ||f(x)_b - f(x')_b ||
\label{eq:contr}
\end{equation}
\item Boundary conditions. For every fundamental region $X$, define the incidence vector $x_X$ whose $a^{\rm th}$ component is 1 if $X \subseteq J_a$ and 0 otherwise. Define $y_X$ likewise, with the $b^{\rm th}$ component decided by whether $X \subseteq K_b$. We demand $f(x_X) = y_X$.  
\end{itemize}

For the second point, we find an RT-realizable vector $s(G)$ such that:
\begin{align}
 \pair{s(G)}{e}&=0 \label{eq:criterion-saturation}\\
 \pair{s(G)}{i_{AB}}&>0 \label{eq:criterion-mutualinfo}
\end{align}
To see that this is sufficient, assume for a contradiction that $e = e' + \textrm{(some subadditivities)}$ and $e'$ is a superbalanced holographic inequality: $\pair{b_{AB}}{e'} = 0$. Then~(\ref{eq:criterion-bell}) requires that among the said subadditivities at least one has $\pair{b_{AB}}{i_{JK}} > 0$, so $A \subseteq J$ and $B \subseteq K$ or vice versa. Focusing on the former option, we combine two instances of strong subadditivity \cite{Lieb:1973cp} with condition~(\ref{eq:criterion-mutualinfo}) to find:
%the monogamy of mutual information~(\ref{eq:mmi})---which $s$ respects because it is RT-realizable---implies that:
\begin{equation}
\pair{s(G)}{i_{JK}} \geq \pair{s(G)}{i_{AK}} \geq \pair{s(G)}{i_{AB}} > 0
\label{eq:mmiapplied}
\end{equation}
Now expand $e$ in equation~(\ref{eq:criterion-saturation}):
\begin{equation}
%\pair{s}{e} = 
\gamma \pair{s(G)}{i_{JK}} + \pair{s(G)}{\,\textrm{other $i_{J'K'}$}} + \pair{s(G)}{e'} = 0
\label{eq:expand}
\end{equation}
Note that the coefficient $\gamma$ of $i_{JK}$ in the expansion must be positive. Inequality~(\ref{eq:mmiapplied}) implies that at least one of the two latter terms is negative. Thus, the entropy vector $s(G)$ is either quantum mechanically inconsistent or not RT-realizable, which yields the desired contradiction.
%This suffices because if $e = i_{XY} + e'$ then (\ref{eq:criterion-saturation}) and (\ref{eq:criterion-mutualinfo}) imply $\pair{s}{e'} < 0$ so $e'$ can be neither a holographic inequality nor a general quantum mechanical inequality like subadditivity. Note that the same argument also excludes other rewritings 

Inequality $e$ might still be written in the form $e' + \textrm{(some subadditivities)}$, but all the subadditivities in question have $\pair{b_{AB}}{i_{J'K'}} = 0$. Therefore, $e'$---the holographic component of $e$---is indeed not superbalanced. Such is the case for inequality~(\ref{eq:witnessineq}) below.
\smallskip

{\it The inequality---} Consider the following inequality $e$:
\begin{equation}
\begin{aligned}
 &5S_{ACDE}+4S_{BCDF}+4S_{BCEG}+5S_{ACFG}\\
 &\quad\geq 4S_B+2S_{ADE}+3S_{ACEF}+2S_{BCDEF}\\
 &\qquad\quad+S_{ACDG}+2S_{AFG}+2S_{BCEFG}.
\end{aligned}
\label{eq:witnessineq}
\end{equation}
%\begin{equation}
%\begin{aligned}
% &5S_{ACDE}+4S_{BCDF}+4S_{BCEG}+5S_{ACFG}\\
% &\quad\geq 4S_B+2S_{ADE}+3S_{ACEF}+2S_{BCDEF}\\
% &\qquad\quad+S_{ACDG}+2S_{AFG}+2S_{BCEFG}.
%\end{aligned}
%\label{eq:witnessineq}
%\end{equation}
%The input bits follow the four LHS regions in their displayed order.
%The output bits follow the RHS regions in their displayed order, with
%an integer coefficient represented by that many unit-weight coordinates.
The contraction map, which establishes that it is valid on all RT-realizable states is:
\begin{table}[h!]
%\caption{The complete weighted binary contraction map for
%Eq.~\eqref{eq:witnessineq}, using the input indices, output order,
%and weights specified in the text. Each input-cube edge in direction
%$a$ has weighted output distance $\alpha_a$.\\}
\label{tab:map}
\centering
\setlength{\tabcolsep}{4pt}
\begin{tabular}{r@{\quad}l@{\qquad}r@{\quad}l}
\toprule
$x$ & $f(x)$ & $x$ & $f(x)$\\
\midrule
0000 & \texttt{0000000000000000} & 0001  & \texttt{0000001100001110}\\
1000 & \texttt{0000111101000000} & 1001  & \texttt{0000111110011100}\\
0100 & \texttt{0000000001100011} & 0101 & \texttt{0000001111101111}\\
1100 & \texttt{0000110001110000} & 1101 & \texttt{0000101111111000}\\
0010 & \texttt{0000000001100011} & 0011 & \texttt{0000000000011111}\\
1010 & \texttt{0000111111100011} & 1011 & \texttt{0000101110011011}\\
0110 & \texttt{1111000001100011} & 0111 & \texttt{0000000001101011}\\
1110 & \texttt{0000100001100011} & 1111 & \texttt{0000001111110011}\\
\bottomrule
\end{tabular}
\end{table}
\newline
The ordering of the bits follows inequality~(\ref{eq:witnessineq}). Note that condition~(\ref{eq:contr}) weighs the bits of $x$ with the coefficients in the inequality. On the other hand, the bits of $f(x)$ all count with weight one; we account for the coefficients on the less-than side of the inequality by repeating bits. One easily verifies that this map satisfies both the contraction condition and all boundary conditions.

\begin{figure}[tb]
\centering
\resizebox{\columnwidth}{!}{%
\begin{tikzpicture}[
  x=1cm,y=1cm,rotate=-90,
  vertex/.style={circle,fill=black,inner sep=0pt,minimum size=1.6mm},
  edge/.style={draw=black!75,line width=.55pt},
  Aedge/.style={draw=blue!65!black,line width=.7pt},
  Bedge/.style={draw=red!65!black,line width=.7pt},
  weight/.style={fill=white,inner sep=1.2pt,font=\small},
  bulklabel/.style={font=\small,inner sep=0pt},
  every label/.style={font=\normalsize,inner sep=1pt}
]
% Rotate coordinates clockwise; node text stays upright.
% Label directions below follow the rotated vertices.
% Regular octagon; v12 divides the two independent bulk cycles.
% w1 and w2 are bulk tree branches. All boundary vertices below are leaves.
\coordinate (u1) at (0.00000000,2.40000000);
\coordinate (v13) at (1.69705627,1.69705627);
\coordinate (u3) at (2.40000000,0.00000000);
\coordinate (v23) at (1.69705627,-1.69705627);
\coordinate (u2) at (0.00000000,-2.40000000);
\coordinate (v24) at (-1.69705627,-1.69705627);
\coordinate (u4) at (-2.40000000,0.00000000);
\coordinate (v14) at (-1.69705627,1.69705627);
\coordinate (v12) at (0.00000000,0.00000000);
\coordinate (w1) at (0.00000000,3.35000000);
\coordinate (w2) at (0.00000000,-3.35000000);

% Distinct boundary vertices may carry the same region label.
\coordinate (A1) at (-0.8,2.7);
\coordinate (A2) at (-0.8,-2.7);
\coordinate (A3) at (3.2,0.45);
\coordinate (A4) at (-3.2,-0.45);
\coordinate (B13) at (2.65,1.85);
\coordinate (B14) at (-2.65,1.85);
\coordinate (B12) at (0.85,0.35);
\coordinate (B23) at (2.65,-1.85);
\coordinate (B24) at (-2.65,-1.85);
\coordinate (C3) at (3.2,-0.45);
\coordinate (C4) at (-3.2,0.45);
\coordinate (D13) at (1.85,2.65);
\coordinate (D1) at (-0.7,3.85);
\coordinate (E14) at (-1.85,2.65);
\coordinate (E1) at (0.7,3.85);
\coordinate (F23) at (1.85,-2.65);
\coordinate (F2) at (0.7,-3.85);
\coordinate (G24) at (-1.85,-2.65);
\coordinate (G2) at (-0.7,-3.85);
\coordinate (O12) at (-0.85,-0.35);
\coordinate (O1) at (0,4.15);
\coordinate (O2) at (0,-4.15);

% All bulk--bulk edges have unit weight.
\draw[edge] (u1)--(v13)--(u3)--(v23)--(u2)--(v24)--(u4)--(v14)--cycle;
\draw[edge] (u1)--(v12)--(u2);
\draw[edge] (u1)--(w1) (u2)--(w2);

% Boundary edges. Only the two marked A edges have weight two.
\draw[Aedge] (u1)--node[weight,pos=.52] {$2$} (A1);
\draw[Aedge] (u2)--node[weight,pos=.52] {$2$} (A2);
\draw[Aedge] (u3)--(A3);
\draw[Aedge] (u4)--(A4);
\draw[Bedge] (v13)--(B13);
\draw[Bedge] (v14)--(B14);
\draw[Bedge] (v12)--(B12);
\draw[Bedge] (v23)--(B23);
\draw[Bedge] (v24)--(B24);
\draw[edge] (u3)--(C3);
\draw[edge] (u4)--(C4);
\draw[edge] (v13)--(D13);
\draw[edge] (w1)--(D1);
\draw[edge] (v14)--(E14);
\draw[edge] (w1)--(E1);
\draw[edge] (v23)--(F23);
\draw[edge] (w2)--(F2);
\draw[edge] (v24)--(G24);
\draw[edge] (w2)--(G2);
\draw[edge] (v12)--(O12);
\draw[edge] (w1)--(O1);
\draw[edge] (w2)--(O2);

% Bulk and boundary vertices use exactly the same solid-dot style.
\node[vertex] at (u1) {};
\node[vertex] at (u2) {};
\node[vertex] at (u3) {};
\node[vertex] at (u4) {};
\node[vertex] at (v13) {};
\node[vertex] at (v14) {};
\node[vertex] at (v23) {};
\node[vertex] at (v24) {};
\node[vertex] at (v12) {};
\node[vertex] at (w1) {};
\node[vertex] at (w2) {};
\node[bulklabel] at (0.3,1.99) {$u_{1}$};
\node[bulklabel] at (0.3,-1.99) {$u_{2}$};
\node[bulklabel] at (1.99,-0.03) {$u_{3}$};
\node[bulklabel] at (-1.99,0.03) {$u_{4}$};
\node[bulklabel] at (1.33,1.33) {$v_{13}$};
\node[bulklabel] at (-1.33,1.33) {$v_{14}$};
\node[bulklabel] at (1.33,-1.33) {$v_{23}$};
\node[bulklabel] at (-1.33,-1.33) {$v_{24}$};
\node[bulklabel] at (-0.31,0.22) {$v_{12}$};
\node[bulklabel] at (0.29,3.31) {$w_{1}$};
\node[bulklabel] at (0.29,-3.31) {$w_{2}$};
\node[vertex,label={above:$A$}] at (A1) {};
\node[vertex,label={above:$A$}] at (A2) {};
\node[vertex,label={below:$A$}] at (A3) {};
\node[vertex,label={above:$A$}] at (A4) {};
\node[vertex,label={below:$B$}] at (B13) {};
\node[vertex,label={above:$B$}] at (B14) {};
\node[vertex,label={below right:$B$}] at (B12) {};
\node[vertex,label={below:$B$}] at (B23) {};
\node[vertex,label={above:$B$}] at (B24) {};
\node[vertex,label={below:$C$}] at (C3) {};
\node[vertex,label={above:$C$}] at (C4) {};
\node[vertex,label={right:$D$}] at (D13) {};
\node[vertex,label={above:$D$}] at (D1) {};
\node[vertex,label={right:$E$}] at (E14) {};
\node[vertex,label={below:$E$}] at (E1) {};
\node[vertex,label={left:$F$}] at (F23) {};
\node[vertex,label={below:$F$}] at (F2) {};
\node[vertex,label={left:$G$}] at (G24) {};
\node[vertex,label={above:$G$}] at (G2) {};
\node[vertex,label={above left:$O$}] at (O12) {};
\node[vertex,label={right:$O$}] at (O1) {};
\node[vertex,label={left:$O$}] at (O2) {};
\end{tikzpicture}%
}
\caption{Graph $G$, whose associated entropy vector $s(G)$ satisfies conditions~(\ref{eq:criterion-saturation}) and (\ref{eq:criterion-mutualinfo}) for inequality~(\ref{eq:witnessineq}). Two marked edges have weight 2; all other edges have weight 1. All boundary regions $A, B, C\!\ldots$ are unions of several disjoint asymptotic boundaries.
%A planar drawing of the graph $G$. Each boundary vertex is a
%leaf; repeated uppercase labels denote the same boundary region.
%Bulk and boundary vertices are all solid dots. The two marked edges
%have weight 2; all other edges have weight 1. 
%Blue and red edges end on $A$ and $B$ leaves, respectively.
}
\label{fig:witness}
\end{figure}
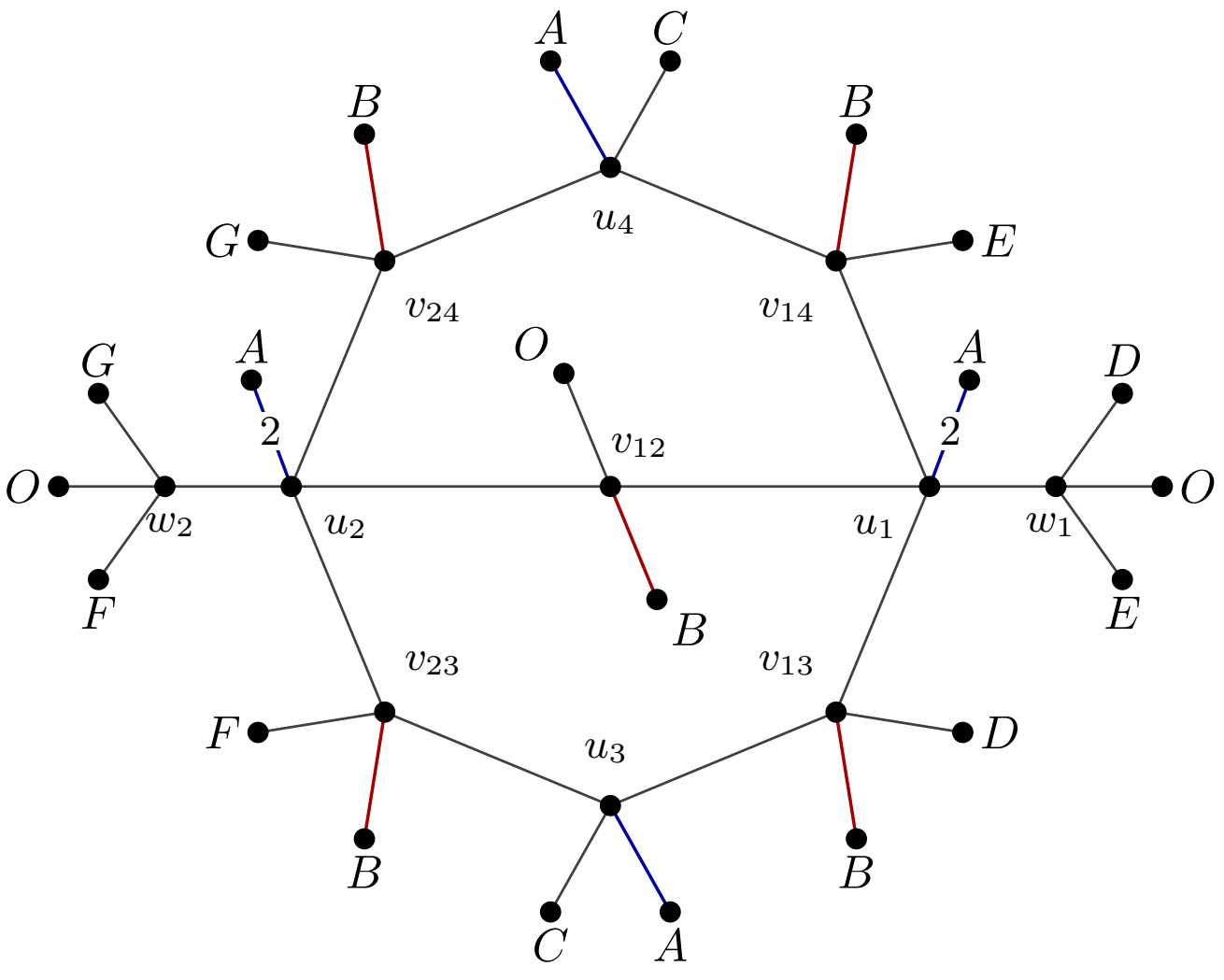

For the RT-realizable vector stipulated in (\ref{eq:criterion-saturation}-\ref{eq:criterion-mutualinfo}), take $s(G)$ derived from graph $G$ in Figure~\ref{fig:witness}. Following an established routine \cite{Bao:2015bfa}, we specify a static slice of a holographic 2+1-dimensional multi-boundary spacetime (and the attendant entropies) using {\bf a graph with weighted edges.} Each {\bf internal} vertex with $l$ incident edges is understood as a topological sphere with $l$ punctures. The {\bf external} vertices represent asymptotic boundaries ($S^1$). The punctures are glued to one another and to the asymptotic boundaries by tubes whose minimal cross-sections are the weights of the graph's edges. With that setup, the entropies of external regions are given by minimal weight cuts through the graph. For more details, see \cite[Sec.~3.2]{Bao:2015bfa}.

%The internal vertices, labelled $u_a$, $v_{ab}$, and $w_a$, with $l$ legs are understood as topological spheres with $l$ punctures. The external vertices are degree-one leaves carrying boundary-region labels and represent asymptotic boundaries. The punctures are glued to one another and to the asymptotic boundaries by tubes whose minimal cross-sections are the weights of the graph's edges. With that setup, the entropies of external regions are given by minimal weight cuts through the graph. For more details, see \cite[Sec.~3.2]{Bao:2015bfa}.

It can be verified by hand that entropy vector $s(G)$ saturates inequality~(\ref{eq:witnessineq}) and that its mutual information is $I(A:B) = 2$. We now introduce the concepts, which enabled us to find $e$ and $s(G)$.
%We found this inequality-graph pair in the course of an extensive dialogue with GPT-6 Astra and GPT-6.1 Sol.
%\smallskip

\section{The dual problem: \\Bell-extractability of mutual information}
\noindent
Graph $G$ has a special property: although $I(A:B) = 2$, we cannot write its entropy vector as a sum of a Bell pair and another RT-realizable entropy vector $s(G')$. Indeed, if $s(G) = b_{AB} + s(G')$ then $\pair{s(G)}{e}=0$ would mean one of two things: either $e$ is superbalanced or, if not, $\pair{s(G')}{e} < 0$ so $s(G')$ is not RT-realizable (or $e$ is not a valid holographic entropy inequality). As we presently explain, this property of graph $G$ is in a rigorous sense dual to the non-superbalance of inequality $e$. 

To state the duality, we introduce two concepts. {\bf Extreme rays} of the holographic entropy cone are RT-realizable entropy vectors, which cannot be written as convex combinations of distinct RT-realizable vectors. Similarly, a {\bf facet} of the holographic entropy cone is an inequality valid on all RT-realizable entropy vectors, which cannot be written as a convex combination of two others. 
We have two useful lemmas:
\begin{itemize}
\item[L1:] If an extreme ray $s(G_p) \neq b_{AB}$ satisfies (\ref{eq:criterion-mutualinfo}) then there exists an inequality $e$ valid on all RT-realizable graphs, which satisfies (\ref{eq:criterion-bell}) and (\ref{eq:criterion-saturation}). It is impossible to rewrite $e = e' + \textrm{(subadditivities)}$ such that $e'$ is also valid on all RT-realizable graphs and $\pair{b_{AB}}{e'} = 0$.
\item[L2:] If a facet $e_q \neq i_{AB}$ satisfies (\ref{eq:criterion-bell}) then there exists an RT-realizable entropy vector $s(G)$, which satisfies (\ref{eq:criterion-saturation}) and (\ref{eq:criterion-mutualinfo}). It is impossible to rewrite $s(G) = s(G') + \textrm{(Bell pairs)}$ such that $s(G')$ is RT-realizable and $\pair{s(G')}{i_{AB}} = 0$.
\end{itemize}
The two lemmas are related by an obvious symmetry. 
We have machine-verified that $s(G)$ from Figure~\ref{fig:witness} is an extreme ray, so inequality~(\ref{eq:witnessineq}) exemplifies L1. By expanding $e$ in facets $e = \sum_p e_p$ we find at least one $e_p$, which exemplifies L2. The claimed existence of $e$ in L1 follows from the extreme ray property of $s(G_p)$: since $s(G_p)$ is RT-realizable but $s(G_p) - \gamma b_{AB}$ for any $\gamma > 0$ is not, there exists an $e$ with $\pair{s(G_p)}{e} = 0$ but $\pair{b_{AB}}{e} > 0$. That $e$ cannot be rewritten as the sum of a superbalanced $e'$ and subadditivities was established below condition~(\ref{eq:criterion-mutualinfo}). L2 is proved by essentially identical arguments.

Thus, non-superbalanced facets are always heralded by RT-realizable states with non-Bell-extractable mutual information. The existence of such states was missed in \cite{He:2020xuo}, which purported to prove that all holographic entropy inequalities are superbalanced. It assumed that every RT-realizable vector $s(G)$ with $I(A:B) = 2\gamma$ can be written as $s(G') + \gamma b_{AB}$ where $s(G')$ is also RT-realizable. Graph $G$ in Figure~\ref{fig:witness} invalidates this assumption.
\smallskip

{\it Tree graphs cannot herald non-superbalanced facets---} That is, an entropy vector $s(T)$ derived from a finite edge-weighted tree graph $T$ cannot be the vector $s(G)$, whose existence is guaranteed by Lemma~L2. The reason is that mutual information borne by tree graphs can always be extracted into Bell pair form in an RT-realizable way: 
\begin{equation}
\exists\, \tilde{T}~\textrm{such that}~s(T) = \tfrac{1}{2} \pair{s(T)}{i_{AB}}\, b_{AB} + s(\tilde{T})
\label{eq:treeclaim}
\end{equation}
This eliminates $s(T)$ because taking $\pair{\ldots}{e_q}$ of both sides of (\ref{eq:treeclaim}) leads to $\pair{s(\tilde{T})}{e_q} < 0$. We prove statement~(\ref{eq:treeclaim}) in Supplemental Material~\cite{Supplement}.

\section{Discussion}
\noindent
Lemma~L1 affords a relatively easy way to find non-superbalanced holographic entropy inequalities: identify extreme rays of the holographic entropy cone with $I(A:B) > 0$. There are at least two reasons why doing so is easier than directly searching for non-superbalanced inequalities. First, an entropy vector derived from a weighted graph does not require any further validation whereas a candidate inequality must be certified by a contraction map. Second, result~(\ref{eq:treeclaim}) imposes an easy-to-implement constraint on candidate graphs: they must contain at least one loop. Although verifying the extremality of an entropy vector is computationally challenging, it is relatively easy to identify high likelihood candidates by testing the realizability of entropy vectors in a neighborhood. Once an extreme ray with non-vanishing mutual information is found, Lemma~L1 guarantees that it saturates at least one non-superbalanced facet of the holographic entropy cone. A wrinkle in this strategy is that the relationship between extreme rays and non-superbalanced facets is not one-to-one: multiple extreme rays with $I(A:B)>0$ can be saturated by a single non-superbalanced facet and vice versa.

We used GPT-6 Astra and GPT-6.1 Sol to search for extreme rays with non-vanishing mutual information. Without attempting an exhaustive search, we found 200,000 such extreme rays; we include their catalogue with the arXiv submission of this paper and exhibit several examples in Supplemental Material~\cite{Supplement}. This count is roughly one order of magnitude below the current total of known extreme rays of the holographic entropy cone \cite{HernandezCuenca:HECDatabase}, which is above 3 million. All our examples come from the holographic entropy cone on $N=7$ named regions; we have not found an $N=6$ example but neither have we proven that one does not exist. It is difficult to ascertain just how common non-superbalanced facets and non-Bell-extractable extreme rays are as a proportion of all facets and extreme rays.
\smallskip

{\it Impact on what we thought we knew---} In light of our results, a number of claims and conjectures about the holographic entropy cone are refuted or require revision:
\begin{itemize}
\item Reference~\cite{Hernandez-Cuenca:2023iqh} conjectured that all facets of the holographic entropy cone are linear combinations of [conditional] tripartite informations. Our results invalidate this conjecture because all such quantities are superbalanced. %For the superbalanced case, see Ref.~\cite{Hubeny:2026vlk}.
\item Reference~\cite{Hernandez-Cuenca:2022pst} conjectured that all extreme rays of the holographic entropy cone can be realized by tree graphs. By decomposing the entropy vector $s(G)$ in Lemma~L2, we obtain a counterexample from every non-superbalanced facet due to result~(\ref{eq:treeclaim}). Graph $G$ in Figure~\ref{fig:witness} is an explicit counterexample.
\item A null reduction of an inequality is obtained by dropping all terms that exclude a given region $X$. If the initial inequality is a superbalanced holographic inequality, its null reduction is also a valid holographic inequality and passes the so-called majorization test; see \cite{Grimaldi:2025jad} for details. We now know that these properties are not universal features of all holographic inequalities. 
\item The majorization test implies two further properties of the initial holographic inequality \cite{Grimaldi:2026lbq, Czech:2026zca}: it has more negative terms than positive terms; the regions in the positive terms have no common overlap. There may yet exist holographic inequalities which do not possess these properties, though we have no examples.
\end{itemize} 

{\it Interpretability---} It would be fascinating to interpret operationally the process of isolating Bell pairs from RT-realizable entropy vectors. Result~(\ref{eq:treeclaim}) suggests that in the AdS$_3$/CFT$_2$ correspondence such an interpretation may relate to the genus of the bulk geometry.

\section*{Acknowledgments}
{\it AI usage statement---}
We used GPT-6 Astra and GPT-6.1 Sol in the computational part of this work. This included the search for extreme rays of the holographic entropy cone with non-vanishing mutual information, finding holographic entropy inequalities which are saturated by the said extreme rays, decomposing them into facets of the holographic entropy cone, and certifying holographic entropy inequalities with contraction maps. The authors take full responsibility for all results and conclusions presented in this paper.

We thank Temple He and Veronika Hubeny for feedback on this manuscript. This work was supported by NSFC Grants NSFC12042505 and NSFC12342501.
The work of XXJ is also supported by the Shuimu Tsinghua Scholar Program of Tsinghua University.
\bibliography{references}
\onecolumngrid
\appendix
\setcounter{secnumdepth}{2}
\section*{Supplemental Material}

\section{Proof that all mutual information derived from tree graph structures is Bell-extractable}
\label{app:bell}
\noindent
Consider a weighted finite tree graph $T$ with $I(A:B) = 2 \gamma > 0$. All edges in $T$ have positive weights. We associate an entropy vector $s(T)$ with this graph, as originally described in Reference~\cite{Bao:2015bfa} and reviewed in the main text. Boundary regions $A$ and $B$ are allowed to comprise multiple external vertices.

We construct another weighted graph $\tilde{T}$ such that $s(\tilde{T}) + \gamma b_{AB} = s(T)$. This proves that tree graphs cannot be the graphs, which are stipulated in Lemma~L2 in the main text.
\smallskip

{\it A special path $\mathcal{P}$ in graph $T$---} 
Consider the `entanglement wedge' $E(AB)$ in $T$, that is the part of $T$ between $AB$ and the minimal cut that computes $S_{AB}$. In case the minimal cut is not unique, we select the $E(AB)$ with the fewest vertices. $I(A:B) > 0$ implies that at least one connected component of $E(AB)$---call it $W$---contains both an $A$-labeled vertex and a $B$-labeled vertex; otherwise the minimal cut for $S_{AB}$ would decompose into an $A$-isolating and a $B$-isolating part and, by minimality, its weight would equal $S_A + S_B$. %The same rationale shows that $E(AB)$ contains fewer connected components than $E(A) \cup E(B)$ does.

In $W$, define the depth $d(v)$ of vertex $v$ to be the number of edges in the unique path from $v$ to some arbitrarily chosen root vertex $r$. Find the deepest vertex $v_0$ with the following property: if we remove the edge that joins $v_0$ to its unique neighbor at depth $d(v_0)-1$, the resulting $v_0$-containing subtree $T(v_0)$ contains both $A$-labeled and $B$-labeled vertices. By construction, all subtrees produced by removing an edge that connects $v_0$ to a neighbor at depth $d(v_0) + 1$ have the opposite property: they cannot contain both $A$-labeled and $B$-labeled vertices. 

Take any path $\mathcal{P}$, which connects an $A$-labeled vertex $a$ to a $B$-labeled vertex $b$ in $T(v_0)$. By the defining property of $T(v_0)$, $\mathcal{P}$ necessarily passes through $v_0$. Path $\mathcal{P}$ obeys the following lemma:
\begin{itemize}
\item[L3:] For every set of external vertices $K$ disjoint from $AB$, the minimal cut that computes $S_{AK}$ contains a single edge in $\mathcal{P}$. 
\end{itemize}

To prove the lemma by contradiction, assume that the minimal $AK$-cut contains three or more edges from $\mathcal{P}$. (That number must be odd so $a$ and $b$ end up on opposite sides of the cut.) Then one or both following circumstances hold:
\begin{enumerate}
\item The minimal cut that computes $AK$ contains two or more edges from the $a$-$v_0$ segment of $\mathcal{P}$.
\item The minimal cut that computes $AK$ contains two or more edges from the $v_0$-$b$ segment of $\mathcal{P}$.
\end{enumerate}

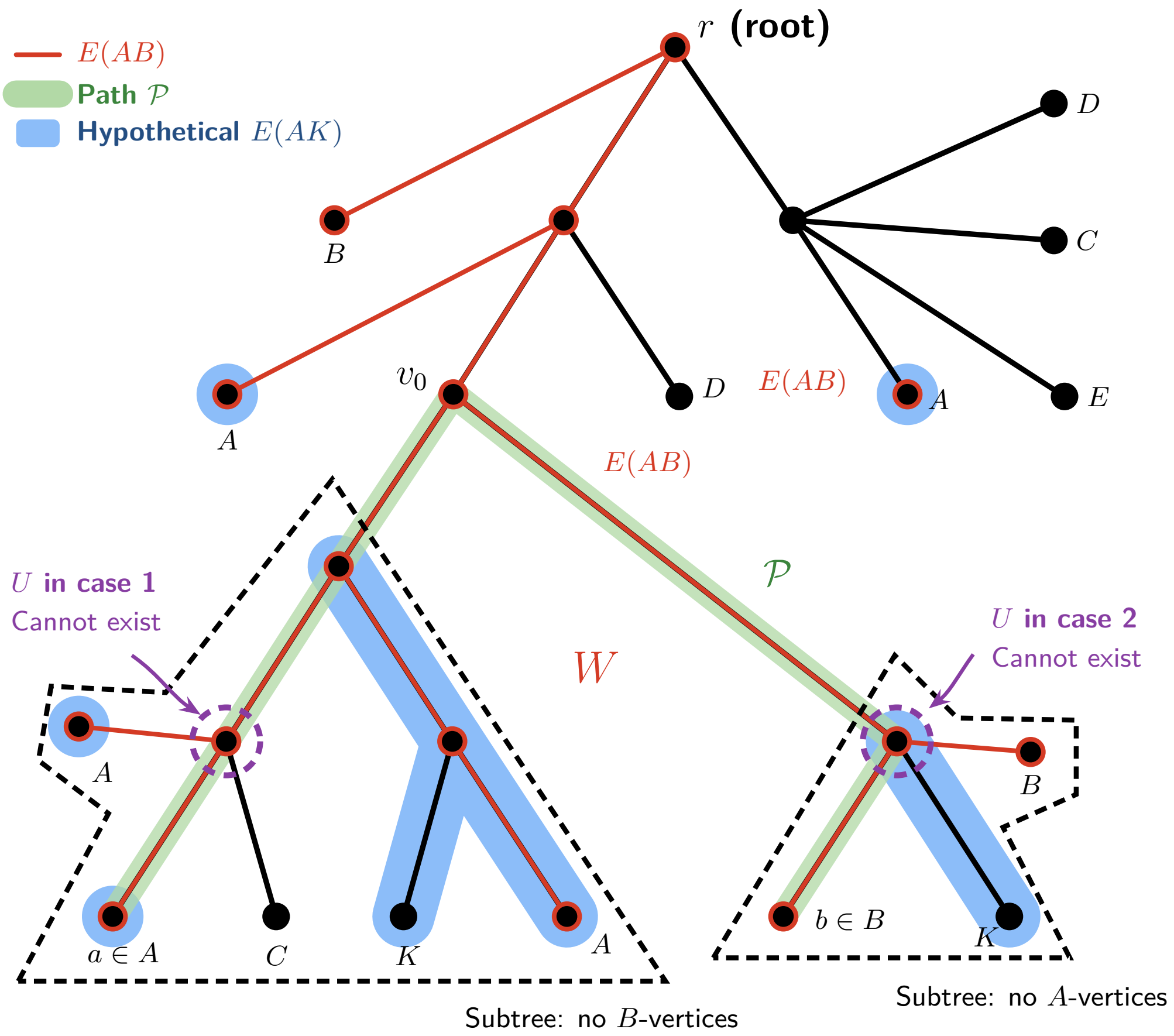
\begin{figure}[tbp]
\centering
\definecolor{EWAB}{HTML}{D43B25}
\definecolor{EWAK}{HTML}{8CBDF9}
\definecolor{EWAKink}{HTML}{244E82}
\definecolor{MainPath}{HTML}{B2D9A6}
\definecolor{PathInk}{HTML}{3C853D}
\definecolor{RegionU}{HTML}{873DA2}
\begin{tikzpicture}[
  x=.012cm,y=-.012cm,
  font=\sffamily,
  tree/.style={black,line width=1.75pt,line cap=round,line join=round},
  red wedge/.style={draw=EWAB,line width=1.5pt,line cap=round,line join=round},
  subtree/.style={black,line width=1.6pt,dash pattern=on 4pt off 2.5pt},
  u outline/.style={draw=RegionU,line width=1.9pt,dash pattern=on 4pt off 3pt},
  u leader/.style={draw=RegionU,line width=1.2pt,-{Stealth[length=2.4mm]}},
  note/.style={inner sep=0pt,outer sep=0pt,anchor=west,align=left},
]
% The coordinates follow the supplied sketch, with y increasing down the page.
\coordinate (r) at (708,106);
\coordinate (rootB) at (387,269);
\coordinate (p) at (603,269);
\coordinate (q) at (819,269);
\coordinate (branchA) at (286,433);
\coordinate (v) at (499,433);
\coordinate (d) at (712,435);
\coordinate (ap) at (927,433);
\coordinate (e) at (1075,435);
\coordinate (extraC) at (1065,288);
\coordinate (extraD) at (1065,159);
\coordinate (l) at (391,595);
\coordinate (u1) at (285,760);
\coordinate (m) at (498,760);
\coordinate (u2) at (917,760);
\coordinate (uLeafA) at (146,746);
\coordinate (uLeafB) at (1043,770);
\coordinate (a) at (178,925);
\coordinate (c) at (332,925);
\coordinate (k1) at (452,925);
\coordinate (aa) at (606,925);
\coordinate (b) at (810,925);
\coordinate (k2) at (1023,925);

% Blue follows the vertices and edges of the hypothetical E(AK).
\fill[EWAK] (a) circle[radius=29] (ap) circle[radius=29] (branchA) circle[radius=29]
  (uLeafA) circle[radius=29];
\draw[EWAK,line width=20pt,line cap=round,line join=round]
  (l)--(m)--(aa) (m)--(k1)
  (u2)--(k2);

% Path P, then the tree and the red E(AB) edges.
\draw[draw=MainPath,line width=8.7pt,opacity=.76,line cap=round,line join=round]
  (a)--(u1)--(l)--(v)--(u2)--(b);
\draw[tree] (r)--(p)--(v)--(l)--(u1)--(a)
  (r)--(q)--(ap) (q)--(e) (q)--(extraC) (q)--(extraD) (p)--(d)
  (v)--(u2)--(b) (u2)--(k2)
  (l)--(m)--(aa) (m)--(k1) (u1)--(c);
\draw[red wedge] (r)--(p)--(v)--(l)--(u1)--(a)
  (v)--(u2)--(b) (l)--(m)--(aa)
  (r)--(rootB) (p)--(branchA) (u1)--(uLeafA) (u2)--(uLeafB);

% The two descendant subtrees and the two choices of U.
\draw[subtree] (89,986)--(174,828)--(108,778)--(120,708)--(228,714)
  --(385,513)--(699,986)--cycle;
\draw[subtree] (744,964)--(915,679)--(974,738)--(1086,740)
  --(1086,810)--(1017,841)--(1080,964)--cycle;
\draw[u outline] (u1) circle[radius=32];
\draw[u outline] (u2) circle[radius=32];
\draw[u leader] (196,679) to[out=-35,in=147] (261,735);
\draw[u leader] (989,678) to[out=-126,in=38] (947,729);

% Draw vertex dots above every region and edge.
\foreach \pt in {q,d,e,extraC,extraD,c,k1,k2}
  \fill[black] (\pt) circle[radius=13];
\foreach \pt in {r,p,v,l,u1,m,u2,a,aa,b,ap,rootB,branchA,uLeafA,uLeafB}
  \filldraw[fill=black,draw=EWAB,line width=1.5pt] (\pt) circle[radius=12];

% Typeset labels in the same parts of the figure as in the sketch.
\node[note,font=\sffamily\bfseries\large] at (729,88) {$r$ (root)};
\node[note,anchor=east,xshift=-8pt,yshift=5pt,font=\sffamily\bfseries\large] at (v) {$v_0$};
\node[note,text=EWAB,font=\sffamily\bfseries] at (640,499) {$E(AB)$};
\node[note,text=EWAB,font=\sffamily\bfseries] at (786,422) {$E(AB)$};
\node[note,text=EWAB,font=\sffamily\bfseries\Large] at (611,690) {$W$};
\node[note,text=PathInk,font=\sffamily\bfseries\large] at (791,601) {$\mathcal P$};

\node[note,anchor=north] at (387,291) {$B$};
\node[note,anchor=north] at (286,467) {$A$};
\node[note] at (733,427) {$D$};
\node[note] at (946,438) {$A$};
\node[note] at (1096,435) {$E$};
\node[note] at (1086,288) {$C$};
\node[note] at (1086,159) {$D$};
\node[note,anchor=north] at (168,780) {$A$};
\node[note,anchor=north] at (1043,792) {$B$};
\node[note] at (154,960) {$a\in A$};
\node[note] at (322,962) {$C$};
\node[note] at (444,962) {$K$};
\node[note] at (628,951) {$A$};
\node[note] at (840,928) {$b\in B$};
\node[note] at (989,944) {$K$};

\node[note,text=RegionU] at (82,628)
  {\textbf{$U$ in case 1}\\[2pt]Cannot exist};
\node[note,text=RegionU] at (1006,662)
  {\textbf{$U$ in case 2}\\[2pt]Cannot exist};
\node[note,font=\sffamily\small] at (510,1020)
  {Subtree: no $B$-vertices};
\node[note,font=\sffamily\small] at (916,1000)
  {Subtree: no $A$-vertices};

% Align the three legend labels with their colour samples.
\draw[red wedge] (87,113)--(128,113);
\node[note,text=EWAB,font=\sffamily\bfseries] at (144,113) {$E(AB)$};
\draw[draw=MainPath,line width=8.7pt,line cap=round] (87,150)--(128,150);
\node[note,text=PathInk,font=\sffamily\bfseries] at (144,150) {Path $\mathcal P$};
\fill[EWAK,rounded corners=2pt] (87,174) rectangle (128,200);
\node[note,text=EWAKink,font=\sffamily\bfseries] at (144,187)
  {Hypothetical $E(AK)$};
\end{tikzpicture}
\caption{The setup of the proof in Appendix~\ref{app:bell}. We display entanglement wedge $E(AB)$, its connected component $W$, the root $r$ and vertex $v_0$, and path $\mathcal{P}$. We also mark a hypothetical cut for $E(AK)$, which cannot be minimal because then excluding regions $U$ from $E(AB)$ would decrease $S_{AB}$.}
\label{fig:tree-bell-proof}
\end{figure}

Remove all edges in the minimal cut for $AK$. This splits $W$ into mutually disjoint components because $\mathcal{P}$ is the only connection from $a$ to $b$. In case 1. above, let $U$ be the connected component that is closest to $a$ along $\mathcal{P}$ but does not contain $a$ (see Fig.~\ref{fig:tree-bell-proof}). This component lives in the subtree of $T(v_0)$, which was constructed to contain no $B$-vertices.  On the other hand, $U$ is separated from $a \in A$ by the minimal cut for $AK$, so it cannot contain external vertices from $AK$. Finally, because $U \subset W \subset E(AB)$, it can only contain external external vertices from $AB$. In summary, $U$ contains no external (boundary) vertices at all. In case 2. above, let $U$ be the connected component that is closest to $b$ but does not contain $b$ (see Fig.~\ref{fig:tree-bell-proof}). By an argument, which is identical except for the exchange $A \leftrightarrow B$ and $X \leftrightarrow \overline{ABK}$, the $U$ chosen in this way also contains no external (boundary) vertices.

We now establish Lemma~L3 by arguing that $U$ must be empty. Because it contains no external vertices, it can be glued to or amputated from any subregion of $T$ without affecting the homology condition in the Ryu-Takayanagi proposal. Let us divide the edges that join $U$ to the rest of $T$ into two classes: those connecting with the rest of $W$ and those leading outside $W$. Because $U \subset E(AB)$, their total weights satisfy:
\begin{equation}
w_{\rm tot} (U, W \setminus U) - w_{\rm tot} (U, T \setminus W) > 0 
\label{eq:edge-comp}
\end{equation}
This is a strict inequality; with equality $U$ could be excluded from $E(AB)$ at no cost---which is impossible because $E(AB)$ contains the least possible number of vertices. But now (\ref{eq:edge-comp}) implies that adjoining $U$ to $E(AK)$ in case 1. (respectively, removing $U$ from $E(AK)$ in case 2.) would decrease the total weight of the $AK$-minimal cut, yielding the desired contradiction.

Because $\mathcal{P} \subset W \subset E(AB) \subset E(ABK)$, we also have:
\begin{itemize}
\item[L4:] For every set of external vertices $K$ disjoint from $AB$, the minimal cut that computes $S_{ABK}$ contains no edges in $\mathcal{P}$. 
\end{itemize}
Lemmas~L3 and L4 mean that the entropy vector $s(T)$ behaves very regularly under changes of edge weights in path $\mathcal{P}$. This regularity allows us to find a new edge-weighted graph $\tilde{T}$ such that $s(\tilde{T}) + \gamma b_{AB} = s(T)$.
\smallskip

\paragraph{Extraction of Bell pairs}
Construct a new graph $T_\tau$ by modifying $T$ in the following way:
\begin{itemize}
\item The weights of all edges not in path $\mathcal{P}$ remain unchanged.
\item The weights of all edges in path $\mathcal{P}$ are uniformly decreased by $\tau$.
\item The weights of all edges are to remain non-negative, with edges of weight-0 considered to be non-existent.
\end{itemize}
Find the largest $t$ with the following property: $S_{AB}$ is the same in $T_t$ as it is in $T$. There are two possible reasons why $t$ cannot be increased any further:
\begin{enumerate}
\item The minimal cut that computes $S_{AB}$ in $T_t$ is at a phase transition. That is, another $AB$-homologous cut---which is not minimal in $T$---becomes degenerate with the cut inherited from $T$. In this case, we automatically know that the new cut intersects path $\mathcal{P}$ two or more times. If it did not intersect $\mathcal{P}$, its weight would be independent of the deformation parameter $\tau$.
\item The weight of one or more edges in $\mathcal{P}$ is set to zero, so we remove an edge (or edges) altogether. In the initial graph $T$, these edges had weight $t$.
\end{enumerate}

Let us relate the entropy vector $s(T_t)$ to $s(T)$. For $0 \leq \tau < t$, no vanishing weight breaks path $\mathcal{P}$ and no phase transition removes it from $E(AB)$, so Lemmas~L3 and L4 continue to apply to it. In particular, L3 tells us that the minimal $AK$-cut in $T_\tau$ is a cut with a single edge in $\mathcal{P}$, its weight lowered by $\tau$. Likewise, L4 says that the minimal $ABK$-cut in $T_\tau$ is a cut with no edges in $\mathcal{P}$, unaffected by $\tau$. Minimizing over all such cuts in $T_\tau$ reproduces the minimal cuts from $T$ and therefore:
\begin{equation}
S_{AK}^{(T_\tau)} = S_{AK}^{(T)} - \tau \qquad {\rm and} \qquad S_{ABK}^{(T_\tau)} = S_{ABK}^{(T)}
\label{eq:intermediate}
\end{equation}
When $\tau = t$, new cuts for $AK$ and $ABK$ may become minimal. However, any new cut that is alternative to the minimal cut inherited from $T$ has a weight greater than or equal to (\ref{eq:intermediate}), for all $0 \leq \tau < t$. Therefore, by continuity, its weight in $T_t$ cannot be strictly smaller than (\ref{eq:intermediate}). Since those values are realized by the cuts inherited from $T$, we conclude that:
\begin{equation}
S_{AK}^{(T_t)} = S_{AK}^{(T)} - t \qquad {\rm and} \qquad S_{ABK}^{(T_t)} = S_{ABK}^{(T)}
\end{equation}
In other words, $s(T_t) + t\, b_{AB} = s(T)$. 

In $T_t$, we find the new $E(AB)$ with the least vertices. We claim that the number of connected components in this new $E(AB)$ is strictly greater than it was in graph $T$. This is automatic if an edge in $\mathcal{P}$ gets erased by setting its weight to 0 because $\mathcal{P}$ is the only connection between the endpoints $a \in A$ and $b \in B$. In the other $\tau$-maximizing scenario---a phase transition---the new minimal $AB$-cut also severs $\mathcal{P}$. At the same time, the weight modification cannot glue together initially disjoint components of $E(AB)$. Doing so would require adjoining a bridging region $U'$ to $E(AB)$, but the edges connecting $U'$ to its complement are unaffected by the weight modification.
\smallskip

{\it Recursive Bell pair extraction exhausts the mutual information---}
We have described a way to trade a weighted tree graph $T$ for a new graph $T_t$ such that $s(T_t) + t\, b_{AB} = s(T)$; the new $T_t$ is a tree or a disjoint union of trees (a forest). Let us refer to this procedure as Bell pair extraction. After one instance of it, apply Bell pair extraction to each connected component of the resulting $T_t$ and continue this process recursively. Bell pair extraction can always be done so long as $E(AB)$ contains at least one connected component with at least one $A$-vertex and at least one $B$-vertex in it. On the other hand, the recursive process cannot continue for infinitely many steps because every application of it increases the number of connected components of $E(AB)$, which in a finite graph is bounded from above by the total number of vertices in $AB$. Therefore, the recursive process must terminate after finitely many steps.

Let $\tilde{T}$ be the weighted graph, which is obtained after the recursive process terminates. The preceding paragraph implies that $E(AB)$ in $\tilde{T}$ contains no connected component with both $A$- and $B$-vertices in it. Therefore, the minimal $AB$-cut in $\tilde{T}$ is the union of disjoint cuts, which are individually homologous to $A$ and $B$. By minimization, this implies that $I(A:B) = 0$ in $\tilde{T}$.

By construction, $s(\tilde{T}) + \tilde{t}\, b_{AB} = s(T)$ for some $\tilde{t}$. Taking $\pair{\ldots}{i_{AB}}$ of both sides, we find $\tilde{t} = \gamma$, as claimed. This concludes the proof.

\section{Non-superbalanced facets and extreme rays with mutual information}
\label{app:examples}
\noindent
The arXiv submission of this paper contains a catalogue of 200,000 extreme rays of the holographic entropy cone with non-vanishing mutual information $I(A:B)$. As explained in the main text, each of them heralds a non-superbalanced facet of the holographic entropy cone, though the relationship is not one-to-one so the count of independent heralded facets may be lower than 200,000. The search for these 200,000 examples was not intended to be exhaustive. All our examples are drawn from the holographic entropy inequality on $N=7$ named regions $A,B,C,D,E,F,G$ (and an unnamed purifier $O$). We have not found an $N=6$ example; neither have we proven that one does not exist.

In this Appendix, we show several explicit pairs $(G,e)$ and evidence that they satisfy relationships~\eqref{eq:criterion-bell}, \eqref{eq:criterion-saturation}, \eqref{eq:criterion-mutualinfo}
in the main text. We further show several examples of non-superbalanced facets. 
%Throughout, \emph{NSNS} denotes non-superbalanced and non-subadditivity. 
As in the main text, $\langle s(G),e\rangle$ means the value of the entropy combination $e$ (LHS$-$RHS) on the entropy vector $s(G)$.

\subsection{Graph realizations of extreme rays}
\label{app:hec7-graphs}
\noindent
We show six relatively uncomplicated graphs $H_r$, which underlie extreme rays of the holographic entropy cone and which have $I(A:B) = 2$. We have machine-verified that each of them saturates $2^7-2 = 126$ linear independent holographic entropy inequalities.

% Here the boundary labels are $\mathcal B_7=\{A,B,C,D,E,F,G,O\}$.
% Among the graph realizations with integer edge weights in our original
% catalogue of 144 distinct extreme-ray orbits, we selected the six with
% the smallest total edge weight $W(G)$
% after excluding the graph already shown in Fig.~\ref{fig:witness}.
% We denote them by $H_1,\ldots,H_6$; their catalogue identifiers are
% $R_{18},R_{22},R_{64},R_1,R_{16},R_{17}$, respectively. The main-text
% graph is $R_0$. In the same order, the six displayed graphs have
% \[
% \begin{aligned}
% W(H_a)&=(31,34,34,36,37,37)_a,\\
% |E(H_a)|&=(30,33,33,34,35,36)_a,
% \end{aligned}
% \]
% where $|E(H_a)|$ denotes the total number of edges in the graph $H_a$.
% All six graphs shown below have $I_{H_a}(A:B)=2$, and each has 126
% linearly independent saturated HEIs (not listed here) certifying extremality.
% \raggedbottom
% \par\smallskip

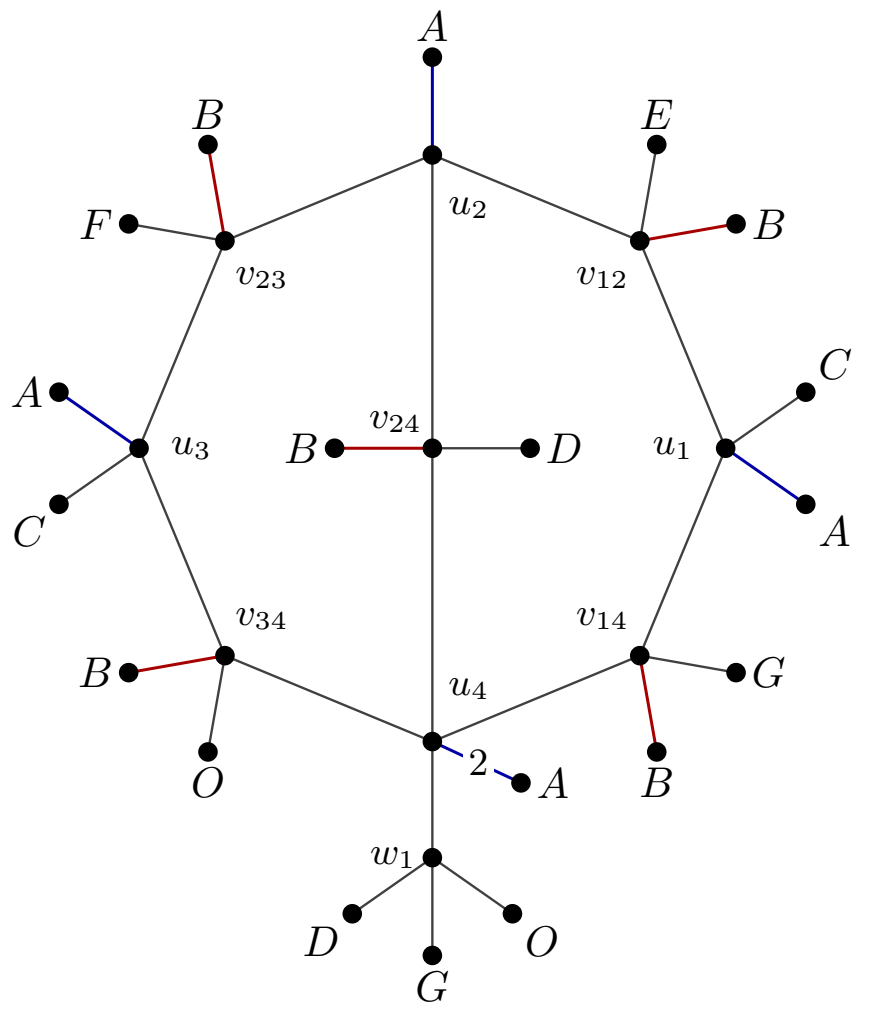
\begin{center}
\begin{minipage}{\linewidth}
\noindent
\begin{minipage}[t]{.32\linewidth}
\centering\vspace{0pt}
\resizebox{.98\linewidth}{!}{%
\begin{tikzpicture}[
  x=1cm,y=1cm,
  vertex/.style={circle,fill=black,inner sep=0pt,minimum size=1.6mm},
  edge/.style={draw=black!75,line width=.55pt},
  Aedge/.style={draw=blue!65!black,line width=.7pt},
  Bedge/.style={draw=red!65!black,line width=.7pt},
  weight/.style={fill=white,inner sep=1.2pt,font=\small},
  bulklabel/.style={font=\small,inner sep=0pt},
  every label/.style={font=\normalsize,inner sep=1pt}
]
% H1: polygonal bulk core and repeated boundary leaves.
\coordinate (n10) at (0.00000000,2.40000000);
\coordinate (n12) at (1.69705627,1.69705627);
\coordinate (n8) at (2.40000000,0.00000000);
\coordinate (n14) at (1.69705627,-1.69705627);
\coordinate (n11) at (0.00000000,-2.40000000);
\coordinate (n15) at (-1.69705627,-1.69705627);
\coordinate (n9) at (-2.40000000,-0.00000000);
\coordinate (n13) at (-1.69705627,1.69705627);
\coordinate (n16) at (0.00000000,0.00000000);
\coordinate (n17) at (0.00000000,-3.35000000);
\coordinate (leaf27) at (-0.65532164,-3.80886115);
\coordinate (leaf28) at (0.00000000,-4.15000000);
\coordinate (leaf29) at (0.65532164,-3.80886115);
\coordinate (leaf0) at (3.05532164,-0.45886115);
\coordinate (leaf20) at (3.05532164,0.45886115);
\coordinate (leaf2) at (0.00000000,3.20000000);
\coordinate (leaf1) at (-3.05532164,0.45886115);
\coordinate (leaf21) at (-3.05532164,-0.45886115);
\coordinate (leaf3) at (0.72504623,-2.73809461);
\coordinate (leaf4) at (2.48490248,1.83597482);
\coordinate (leaf22) at (1.83597482,2.48490248);
\coordinate (leaf6) at (1.83597482,-2.48490248);
\coordinate (leaf24) at (2.48490248,-1.83597482);
\coordinate (leaf5) at (-1.83597482,2.48490248);
\coordinate (leaf23) at (-2.48490248,1.83597482);
\coordinate (leaf8) at (-0.80000000,0.00000000);
\coordinate (leaf26) at (0.80000000,-0.00000000);
\coordinate (leaf7) at (-2.48490248,-1.83597482);
\coordinate (leaf25) at (-1.83597482,-2.48490248);
\draw[edge] (n8)--(n12);
\draw[edge] (n8)--(n14);
\draw[edge] (n9)--(n13);
\draw[edge] (n9)--(n15);
\draw[edge] (n10)--(n12);
\draw[edge] (n10)--(n13);
\draw[edge] (n10)--(n16);
\draw[edge] (n11)--(n14);
\draw[edge] (n11)--(n15);
\draw[edge] (n11)--(n16);
\draw[edge] (n11)--(n17);
\draw[Aedge] (n8)--(leaf0);
\draw[Aedge] (n9)--(leaf1);
\draw[Aedge] (n10)--(leaf2);
\draw[Aedge] (n11)--node[weight,pos=.52] {$ 2 $} (leaf3);
\draw[Bedge] (n12)--(leaf4);
\draw[Bedge] (n13)--(leaf5);
\draw[Bedge] (n14)--(leaf6);
\draw[Bedge] (n15)--(leaf7);
\draw[Bedge] (n16)--(leaf8);
\draw[edge] (n8)--(leaf20);
\draw[edge] (n9)--(leaf21);
\draw[edge] (n12)--(leaf22);
\draw[edge] (n13)--(leaf23);
\draw[edge] (n14)--(leaf24);
\draw[edge] (n15)--(leaf25);
\draw[edge] (n16)--(leaf26);
\draw[edge] (n17)--(leaf27);
\draw[edge] (n17)--(leaf28);
\draw[edge] (n17)--(leaf29);
\node[vertex] at (n17) {};
\node[bulklabel] at (-0.32000000,-3.35000000) {$ w_{1} $};
\node[vertex] at (n8) {};
\node[bulklabel] at (1.97000000,0.00000000) {$ u_{1} $};
\node[vertex] at (n10) {};
\node[bulklabel] at (0.30000000,1.97000000) {$ u_{2} $};
\node[vertex] at (n9) {};
\node[bulklabel] at (-1.97000000,-0.00000000) {$ u_{3} $};
\node[vertex] at (n11) {};
\node[bulklabel] at (0.30000000,-1.97000000) {$ u_{4} $};
\node[vertex] at (n12) {};
\node[bulklabel] at (1.39300036,1.39300036) {$ v_{12} $};
\node[vertex] at (n14) {};
\node[bulklabel] at (1.39300036,-1.39300036) {$ v_{14} $};
\node[vertex] at (n13) {};
\node[bulklabel] at (-1.39300036,1.39300036) {$ v_{23} $};
\node[vertex] at (n16) {};
\node[bulklabel] at (-0.30000000,0.22000000) {$ v_{24} $};
\node[vertex] at (n15) {};
\node[bulklabel] at (-1.39300036,-1.39300036) {$ v_{34} $};
\node[vertex,label={below right:$A$}] at (leaf0) {};
\node[vertex,label={left:$A$}] at (leaf1) {};
\node[vertex,label={above:$A$}] at (leaf2) {};
\node[vertex,label={right:$A$}] at (leaf3) {};
\node[vertex,label={right:$B$}] at (leaf4) {};
\node[vertex,label={above:$B$}] at (leaf5) {};
\node[vertex,label={below:$B$}] at (leaf6) {};
\node[vertex,label={left:$B$}] at (leaf7) {};
\node[vertex,label={left:$B$}] at (leaf8) {};
\node[vertex,label={above right:$C$}] at (leaf20) {};
\node[vertex,label={below left:$C$}] at (leaf21) {};
\node[vertex,label={above:$E$}] at (leaf22) {};
\node[vertex,label={left:$F$}] at (leaf23) {};
\node[vertex,label={right:$G$}] at (leaf24) {};
\node[vertex,label={below:$O$}] at (leaf25) {};
\node[vertex,label={right:$D$}] at (leaf26) {};
\node[vertex,label={below left:$D$}] at (leaf27) {};
\node[vertex,label={below:$G$}] at (leaf28) {};
\node[vertex,label={below right:$O$}] at (leaf29) {};
\end{tikzpicture}%
}
\par\vspace{2pt}$H_{1}$
\end{minipage}\hfill%
\begin{minipage}[t]{.32\linewidth}
\centering\vspace{0pt}
\resizebox{.98\linewidth}{!}
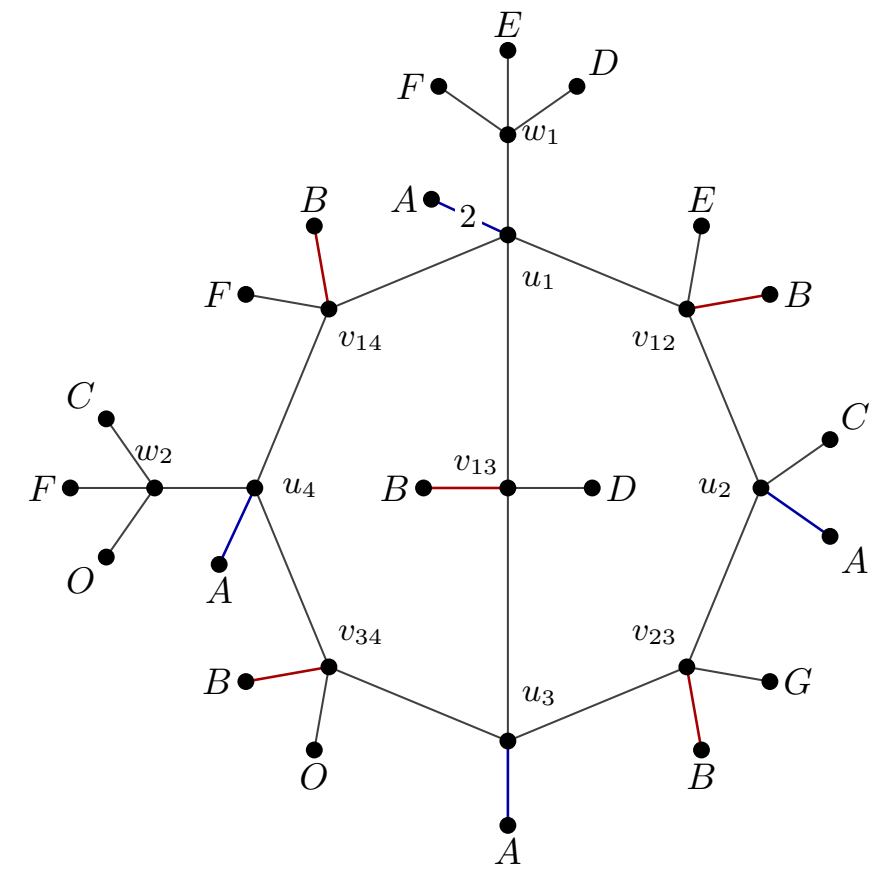
{%
\begin{tikzpicture}[
  x=1cm,y=1cm,
  vertex/.style={circle,fill=black,inner sep=0pt,minimum size=1.6mm},
  edge/.style={draw=black!75,line width=.55pt},
  Aedge/.style={draw=blue!65!black,line width=.7pt},
  Bedge/.style={draw=red!65!black,line width=.7pt},
  weight/.style={fill=white,inner sep=1.2pt,font=\small},
  bulklabel/.style={font=\small,inner sep=0pt},
  every label/.style={font=\normalsize,inner sep=1pt}
]
% H2: polygonal bulk core and repeated boundary leaves.
\coordinate (n9) at (0.00000000,2.40000000);
\coordinate (n13) at (1.69705627,1.69705627);
\coordinate (n11) at (2.40000000,0.00000000);
\coordinate (n14) at (1.69705627,-1.69705627);
\coordinate (n10) at (0.00000000,-2.40000000);
\coordinate (n15) at (-1.69705627,-1.69705627);
\coordinate (n12) at (-2.40000000,-0.00000000);
\coordinate (n17) at (-1.69705627,1.69705627);
\coordinate (n16) at (0.00000000,0.00000000);
\coordinate (n18) at (-3.35000000,-0.00000000);
\coordinate (n8) at (0.00000000,3.35000000);
\coordinate (leaf21) at (0.65532164,3.80886115);
\coordinate (leaf22) at (0.00000000,4.15000000);
\coordinate (leaf23) at (-0.65532164,3.80886115);
\coordinate (leaf30) at (-3.80886115,0.65532164);
\coordinate (leaf31) at (-4.15000000,-0.00000000);
\coordinate (leaf32) at (-3.80886115,-0.65532164);
\coordinate (leaf0) at (-0.72504623,2.73809461);
\coordinate (leaf2) at (3.05532164,-0.45886115);
\coordinate (leaf24) at (3.05532164,0.45886115);
\coordinate (leaf1) at (0.00000000,-3.20000000);
\coordinate (leaf3) at (-2.73809461,-0.72504623);
\coordinate (leaf4) at (2.48490248,1.83597482);
\coordinate (leaf25) at (1.83597482,2.48490248);
\coordinate (leaf7) at (-0.80000000,0.00000000);
\coordinate (leaf28) at (0.80000000,-0.00000000);
\coordinate (leaf5) at (1.83597482,-2.48490248);
\coordinate (leaf26) at (2.48490248,-1.83597482);
\coordinate (leaf8) at (-1.83597482,2.48490248);
\coordinate (leaf29) at (-2.48490248,1.83597482);
\coordinate (leaf6) at (-2.48490248,-1.83597482);
\coordinate (leaf27) at (-1.83597482,-2.48490248);
\draw[edge] (n8)--(n9);
\draw[edge] (n9)--(n13);
\draw[edge] (n9)--(n16);
\draw[edge] (n9)--(n17);
\draw[edge] (n10)--(n14);
\draw[edge] (n10)--(n15);
\draw[edge] (n10)--(n16);
\draw[edge] (n11)--(n13);
\draw[edge] (n11)--(n14);
\draw[edge] (n12)--(n15);
\draw[edge] (n12)--(n17);
\draw[edge] (n12)--(n18);
\draw[Aedge] (n9)--node[weight,pos=.52] {$ 2 $} (leaf0);
\draw[Aedge] (n10)--(leaf1);
\draw[Aedge] (n11)--(leaf2);
\draw[Aedge] (n12)--(leaf3);
\draw[Bedge] (n13)--(leaf4);
\draw[Bedge] (n14)--(leaf5);
\draw[Bedge] (n15)--(leaf6);
\draw[Bedge] (n16)--(leaf7);
\draw[Bedge] (n17)--(leaf8);
\draw[edge] (n8)--(leaf21);
\draw[edge] (n8)--(leaf22);
\draw[edge] (n8)--(leaf23);
\draw[edge] (n11)--(leaf24);
\draw[edge] (n13)--(leaf25);
\draw[edge] (n14)--(leaf26);
\draw[edge] (n15)--(leaf27);
\draw[edge] (n16)--(leaf28);
\draw[edge] (n17)--(leaf29);
\draw[edge] (n18)--(leaf30);
\draw[edge] (n18)--(leaf31);
\draw[edge] (n18)--(leaf32);
\node[vertex] at (n8) {};
\node[bulklabel] at (0.32000000,3.35000000) {$ w_{1} $};
\node[vertex] at (n18) {};
\node[bulklabel] at (-3.35000000,0.32000000) {$ w_{2} $};
\node[vertex] at (n9) {};
\node[bulklabel] at (0.30000000,1.97000000) {$ u_{1} $};
\node[vertex] at (n11) {};
\node[bulklabel] at (1.97000000,0.00000000) {$ u_{2} $};
\node[vertex] at (n10) {};
\node[bulklabel] at (0.30000000,-1.97000000) {$ u_{3} $};
\node[vertex] at (n12) {};
\node[bulklabel] at (-1.97000000,-0.00000000) {$ u_{4} $};
\node[vertex] at (n13) {};
\node[bulklabel] at (1.39300036,1.39300036) {$ v_{12} $};
\node[vertex] at (n16) {};
\node[bulklabel] at (-0.30000000,0.22000000) {$ v_{13} $};
\node[vertex] at (n14) {};
\node[bulklabel] at (1.39300036,-1.39300036) {$ v_{23} $};
\node[vertex] at (n17) {};
\node[bulklabel] at (-1.39300036,1.39300036) {$ v_{14} $};
\node[vertex] at (n15) {};
\node[bulklabel] at (-1.39300036,-1.39300036) {$ v_{34} $};
\node[vertex,label={left:$A$}] at (leaf0) {};
\node[vertex,label={below:$A$}] at (leaf1) {};
\node[vertex,label={below right:$A$}] at (leaf2) {};
\node[vertex,label={below:$A$}] at (leaf3) {};
\node[vertex,label={right:$B$}] at (leaf4) {};
\node[vertex,label={below:$B$}] at (leaf5) {};
\node[vertex,label={left:$B$}] at (leaf6) {};
\node[vertex,label={left:$B$}] at (leaf7) {};
\node[vertex,label={above:$B$}] at (leaf8) {};
\node[vertex,label={above right:$D$}] at (leaf21) {};
\node[vertex,label={above:$E$}] at (leaf22) {};
\node[vertex,label={left:$F$}] at (leaf23) {};
\node[vertex,label={above right:$C$}] at (leaf24) {};
\node[vertex,label={above:$E$}] at (leaf25) {};
\node[vertex,label={right:$G$}] at (leaf26) {};
\node[vertex,label={below:$O$}] at (leaf27) {};
\node[vertex,label={right:$D$}] at (leaf28) {};
\node[vertex,label={left:$F$}] at (leaf29) {};
\node[vertex,label={above left:$C$}] at (leaf30) {};
\node[vertex,label={left:$F$}] at (leaf31) {};
\node[vertex,label={below left:$O$}] at (leaf32) {};
\end{tikzpicture}%
}
\par\vspace{2pt}$H_{2}$
\end{minipage}\hfill%
\begin{minipage}[t]{.32\linewidth}
\centering\vspace{0pt}
\resizebox{.98\linewidth}{!}
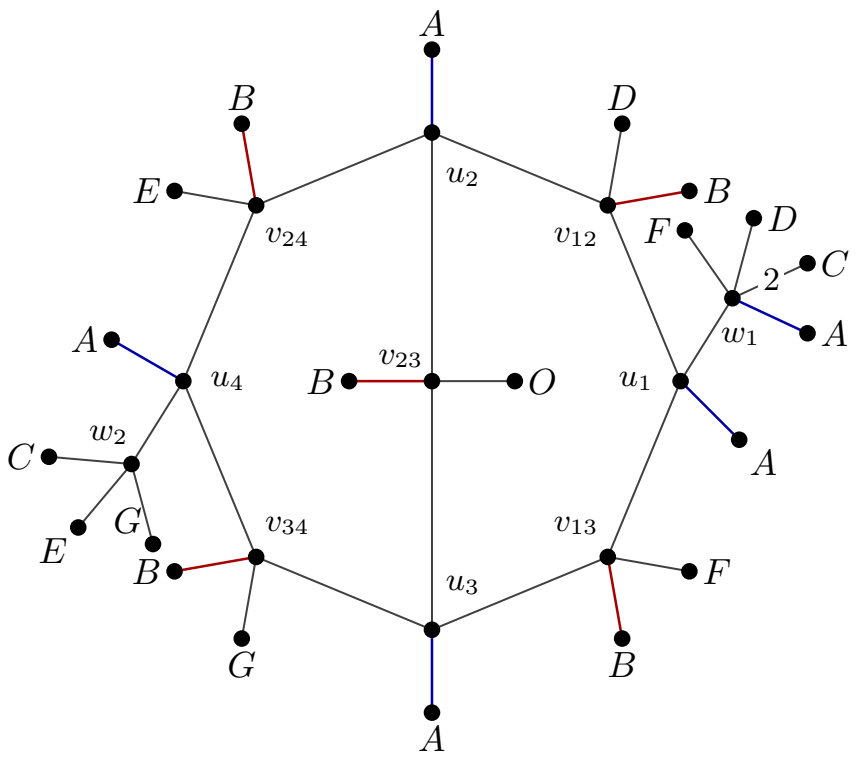
{%
\begin{tikzpicture}[
  x=1cm,y=1cm,
  vertex/.style={circle,fill=black,inner sep=0pt,minimum size=1.6mm},
  edge/.style={draw=black!75,line width=.55pt},
  Aedge/.style={draw=blue!65!black,line width=.7pt},
  Bedge/.style={draw=red!65!black,line width=.7pt},
  weight/.style={fill=white,inner sep=1.2pt,font=\small},
  bulklabel/.style={font=\small,inner sep=0pt},
  every label/.style={font=\normalsize,inner sep=1pt}
]
% H3: polygonal bulk core and repeated boundary leaves.
\coordinate (n11) at (0.00000000,2.40000000);
\coordinate (n12) at (1.69705627,1.69705627);
\coordinate (n16) at (2.40000000,0.00000000);
\coordinate (n15) at (1.69705627,-1.69705627);
\coordinate (n14) at (0.00000000,-2.40000000);
\coordinate (n17) at (-1.69705627,-1.69705627);
\coordinate (n18) at (-2.40000000,-0.00000000);
\coordinate (n13) at (-1.69705627,1.69705627);
\coordinate (n10) at (0.00000000,0.00000000);
\coordinate (n9) at (2.90000000,0.80000000);
\coordinate (n8) at (-2.90000000,-0.80000000);
\coordinate (leaf0) at (3.62504623,0.46190539);
\coordinate (leaf25) at (3.62504623,1.13809461);
\coordinate (leaf26) at (3.10705524,1.57274066);
\coordinate (leaf27) at (2.44113885,1.45532164);
\coordinate (leaf22) at (-3.69695576,-0.73027541);
\coordinate (leaf23) at (-3.41423009,-1.41283555);
\coordinate (leaf24) at (-2.69294476,-1.57274066);
\coordinate (leaf3) at (2.96568542,-0.56568542);
\coordinate (leaf1) at (0.00000000,3.20000000);
\coordinate (leaf2) at (0.00000000,-3.20000000);
\coordinate (leaf4) at (-3.09282032,0.40000000);
\coordinate (leaf6) at (2.48490248,1.83597482);
\coordinate (leaf29) at (1.83597482,2.48490248);
\coordinate (leaf8) at (1.83597482,-2.48490248);
\coordinate (leaf31) at (2.48490248,-1.83597482);
\coordinate (leaf5) at (-0.80000000,0.00000000);
\coordinate (leaf28) at (0.80000000,-0.00000000);
\coordinate (leaf7) at (-1.83597482,2.48490248);
\coordinate (leaf30) at (-2.48490248,1.83597482);
\coordinate (leaf9) at (-2.48490248,-1.83597482);
\coordinate (leaf32) at (-1.83597482,-2.48490248);
\draw[edge] (n8)--(n18);
\draw[edge] (n9)--(n16);
\draw[edge] (n10)--(n11);
\draw[edge] (n10)--(n14);
\draw[edge] (n11)--(n12);
\draw[edge] (n11)--(n13);
\draw[edge] (n12)--(n16);
\draw[edge] (n13)--(n18);
\draw[edge] (n14)--(n15);
\draw[edge] (n14)--(n17);
\draw[edge] (n15)--(n16);
\draw[edge] (n17)--(n18);
\draw[Aedge] (n9)--(leaf0);
\draw[Aedge] (n11)--(leaf1);
\draw[Aedge] (n14)--(leaf2);
\draw[Aedge] (n16)--(leaf3);
\draw[Aedge] (n18)--(leaf4);
\draw[Bedge] (n10)--(leaf5);
\draw[Bedge] (n12)--(leaf6);
\draw[Bedge] (n13)--(leaf7);
\draw[Bedge] (n15)--(leaf8);
\draw[Bedge] (n17)--(leaf9);
\draw[edge] (n8)--(leaf22);
\draw[edge] (n8)--(leaf23);
\draw[edge] (n8)--(leaf24);
\draw[edge] (n9)--node[weight,pos=.52] {$ 2 $} (leaf25);
\draw[edge] (n9)--(leaf26);
\draw[edge] (n9)--(leaf27);
\draw[edge] (n10)--(leaf28);
\draw[edge] (n12)--(leaf29);
\draw[edge] (n13)--(leaf30);
\draw[edge] (n15)--(leaf31);
\draw[edge] (n17)--(leaf32);
\node[vertex] at (n9) {};
\node[bulklabel] at (2.98000000,0.43000000) {$ w_{1} $};
\node[vertex] at (n8) {};
\node[bulklabel] at (-3.12000000,-0.52000000) {$ w_{2} $};
\node[vertex] at (n16) {};
\node[bulklabel] at (1.97000000,0.00000000) {$ u_{1} $};
\node[vertex] at (n11) {};
\node[bulklabel] at (0.30000000,1.97000000) {$ u_{2} $};
\node[vertex] at (n14) {};
\node[bulklabel] at (0.30000000,-1.97000000) {$ u_{3} $};
\node[vertex] at (n18) {};
\node[bulklabel] at (-1.97000000,-0.00000000) {$ u_{4} $};
\node[vertex] at (n12) {};
\node[bulklabel] at (1.39300036,1.39300036) {$ v_{12} $};
\node[vertex] at (n15) {};
\node[bulklabel] at (1.39300036,-1.39300036) {$ v_{13} $};
\node[vertex] at (n10) {};
\node[bulklabel] at (-0.30000000,0.22000000) {$ v_{23} $};
\node[vertex] at (n13) {};
\node[bulklabel] at (-1.39300036,1.39300036) {$ v_{24} $};
\node[vertex] at (n17) {};
\node[bulklabel] at (-1.39300036,-1.39300036) {$ v_{34} $};
\node[vertex,label={right:$A$}] at (leaf0) {};
\node[vertex,label={above:$A$}] at (leaf1) {};
\node[vertex,label={below:$A$}] at (leaf2) {};
\node[vertex,label={below right:$A$}] at (leaf3) {};
\node[vertex,label={left:$A$}] at (leaf4) {};
\node[vertex,label={left:$B$}] at (leaf5) {};
\node[vertex,label={right:$B$}] at (leaf6) {};
\node[vertex,label={above:$B$}] at (leaf7) {};
\node[vertex,label={below:$B$}] at (leaf8) {};
\node[vertex,label={left:$B$}] at (leaf9) {};
\node[vertex,label={left:$C$}] at (leaf22) {};
\node[vertex,label={below left:$E$}] at (leaf23) {};
\node[vertex,label={above left:$G$}] at (leaf24) {};
\node[vertex,label={right:$C$}] at (leaf25) {};
\node[vertex,label={right:$D$}] at (leaf26) {};
\node[vertex,label={left:$F$}] at (leaf27) {};
\node[vertex,label={right:$O$}] at (leaf28) {};
\node[vertex,label={above:$D$}] at (leaf29) {};
\node[vertex,label={left:$E$}] at (leaf30) {};
\node[vertex,label={right:$F$}] at (leaf31) {};
\node[vertex,label={below:$G$}] at (leaf32) {};
\end{tikzpicture}%
}
\par\vspace{2pt}$H_{3}$
\end{minipage}\par\vspace{7pt}
\begin{minipage}[t]{.32\linewidth}
\centering\vspace{0pt}
\resizebox{.98\linewidth}{!}
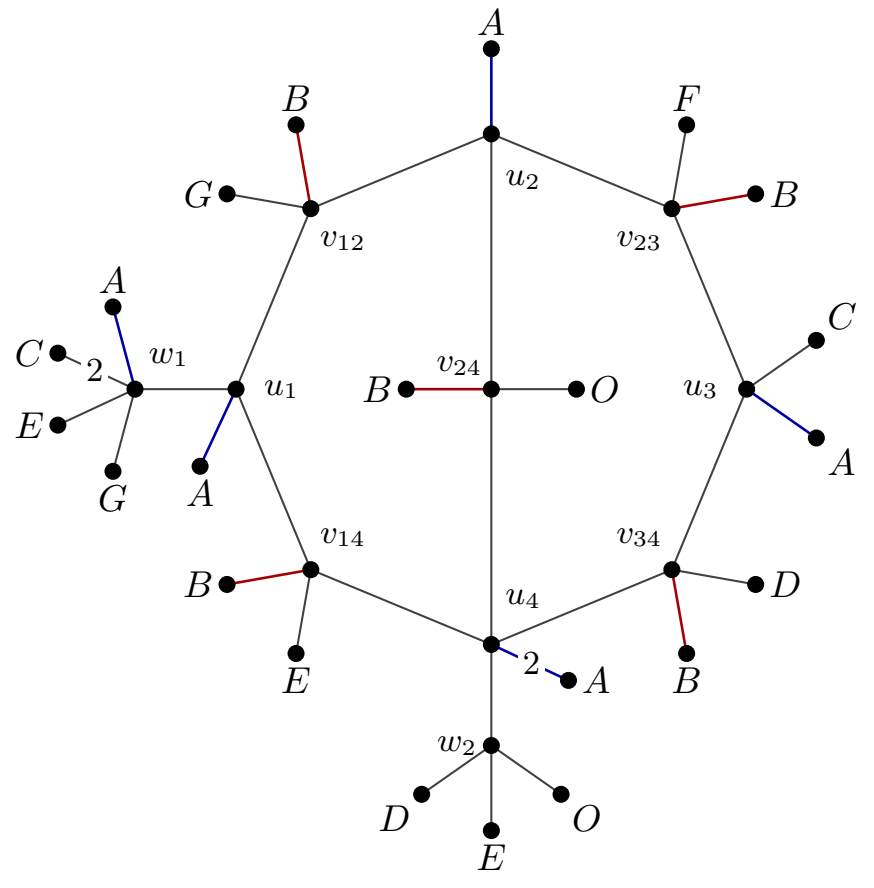
{%
\begin{tikzpicture}[
  x=1cm,y=1cm,
  vertex/.style={circle,fill=black,inner sep=0pt,minimum size=1.6mm},
  edge/.style={draw=black!75,line width=.55pt},
  Aedge/.style={draw=blue!65!black,line width=.7pt},
  Bedge/.style={draw=red!65!black,line width=.7pt},
  weight/.style={fill=white,inner sep=1.2pt,font=\small},
  bulklabel/.style={font=\small,inner sep=0pt},
  every label/.style={font=\normalsize,inner sep=1pt}
]
% H4: polygonal bulk core and repeated boundary leaves.
\coordinate (n11) at (0.00000000,2.40000000);
\coordinate (n15) at (1.69705627,1.69705627);
\coordinate (n8) at (2.40000000,0.00000000);
\coordinate (n13) at (1.69705627,-1.69705627);
\coordinate (n10) at (0.00000000,-2.40000000);
\coordinate (n14) at (-1.69705627,-1.69705627);
\coordinate (n12) at (-2.40000000,-0.00000000);
\coordinate (n16) at (-1.69705627,1.69705627);
\coordinate (n17) at (0.00000000,0.00000000);
\coordinate (n9) at (-3.35000000,-0.00000000);
\coordinate (n18) at (0.00000000,-3.35000000);
\coordinate (leaf1) at (-3.55705524,0.77274066);
\coordinate (leaf23) at (-4.07504623,0.33809461);
\coordinate (leaf24) at (-4.07504623,-0.33809461);
\coordinate (leaf25) at (-3.55705524,-0.77274066);
\coordinate (leaf31) at (-0.65532164,-3.80886115);
\coordinate (leaf32) at (0.00000000,-4.15000000);
\coordinate (leaf33) at (0.65532164,-3.80886115);
\coordinate (leaf4) at (-2.73809461,-0.72504623);
\coordinate (leaf3) at (0.00000000,3.20000000);
\coordinate (leaf0) at (3.05532164,-0.45886115);
\coordinate (leaf22) at (3.05532164,0.45886115);
\coordinate (leaf2) at (0.72504623,-2.73809461);
\coordinate (leaf8) at (-1.83597482,2.48490248);
\coordinate (leaf29) at (-2.48490248,1.83597482);
\coordinate (leaf6) at (-2.48490248,-1.83597482);
\coordinate (leaf27) at (-1.83597482,-2.48490248);
\coordinate (leaf7) at (2.48490248,1.83597482);
\coordinate (leaf28) at (1.83597482,2.48490248);
\coordinate (leaf9) at (-0.80000000,0.00000000);
\coordinate (leaf30) at (0.80000000,-0.00000000);
\coordinate (leaf5) at (1.83597482,-2.48490248);
\coordinate (leaf26) at (2.48490248,-1.83597482);
\draw[edge] (n8)--(n13);
\draw[edge] (n8)--(n15);
\draw[edge] (n9)--(n12);
\draw[edge] (n10)--(n13);
\draw[edge] (n10)--(n14);
\draw[edge] (n10)--(n17);
\draw[edge] (n10)--(n18);
\draw[edge] (n11)--(n15);
\draw[edge] (n11)--(n16);
\draw[edge] (n11)--(n17);
\draw[edge] (n12)--(n14);
\draw[edge] (n12)--(n16);
\draw[Aedge] (n8)--(leaf0);
\draw[Aedge] (n9)--(leaf1);
\draw[Aedge] (n10)--node[weight,pos=.52] {$ 2 $} (leaf2);
\draw[Aedge] (n11)--(leaf3);
\draw[Aedge] (n12)--(leaf4);
\draw[Bedge] (n13)--(leaf5);
\draw[Bedge] (n14)--(leaf6);
\draw[Bedge] (n15)--(leaf7);
\draw[Bedge] (n16)--(leaf8);
\draw[Bedge] (n17)--(leaf9);
\draw[edge] (n8)--(leaf22);
\draw[edge] (n9)--node[weight,pos=.52] {$ 2 $} (leaf23);
\draw[edge] (n9)--(leaf24);
\draw[edge] (n9)--(leaf25);
\draw[edge] (n13)--(leaf26);
\draw[edge] (n14)--(leaf27);
\draw[edge] (n15)--(leaf28);
\draw[edge] (n16)--(leaf29);
\draw[edge] (n17)--(leaf30);
\draw[edge] (n18)--(leaf31);
\draw[edge] (n18)--(leaf32);
\draw[edge] (n18)--(leaf33);
\node[vertex] at (n9) {};
\node[bulklabel] at (-3.03000000,0.30000000) {$ w_{1} $};
\node[vertex] at (n18) {};
\node[bulklabel] at (-0.32000000,-3.35000000) {$ w_{2} $};
\node[vertex] at (n12) {};
\node[bulklabel] at (-1.97000000,-0.00000000) {$ u_{1} $};
\node[vertex] at (n11) {};
\node[bulklabel] at (0.30000000,1.97000000) {$ u_{2} $};
\node[vertex] at (n8) {};
\node[bulklabel] at (1.97000000,0.00000000) {$ u_{3} $};
\node[vertex] at (n10) {};
\node[bulklabel] at (0.30000000,-1.97000000) {$ u_{4} $};
\node[vertex] at (n16) {};
\node[bulklabel] at (-1.39300036,1.39300036) {$ v_{12} $};
\node[vertex] at (n14) {};
\node[bulklabel] at (-1.39300036,-1.39300036) {$ v_{14} $};
\node[vertex] at (n15) {};
\node[bulklabel] at (1.39300036,1.39300036) {$ v_{23} $};
\node[vertex] at (n17) {};
\node[bulklabel] at (-0.30000000,0.22000000) {$ v_{24} $};
\node[vertex] at (n13) {};
\node[bulklabel] at (1.39300036,-1.39300036) {$ v_{34} $};
\node[vertex,label={below right:$A$}] at (leaf0) {};
\node[vertex,label={above:$A$}] at (leaf1) {};
\node[vertex,label={right:$A$}] at (leaf2) {};
\node[vertex,label={above:$A$}] at (leaf3) {};
\node[vertex,label={below:$A$}] at (leaf4) {};
\node[vertex,label={below:$B$}] at (leaf5) {};
\node[vertex,label={left:$B$}] at (leaf6) {};
\node[vertex,label={right:$B$}] at (leaf7) {};
\node[vertex,label={above:$B$}] at (leaf8) {};
\node[vertex,label={left:$B$}] at (leaf9) {};
\node[vertex,label={above right:$C$}] at (leaf22) {};
\node[vertex,label={left:$C$}] at (leaf23) {};
\node[vertex,label={left:$E$}] at (leaf24) {};
\node[vertex,label={below:$G$}] at (leaf25) {};
\node[vertex,label={right:$D$}] at (leaf26) {};
\node[vertex,label={below:$E$}] at (leaf27) {};
\node[vertex,label={above:$F$}] at (leaf28) {};
\node[vertex,label={left:$G$}] at (leaf29) {};
\node[vertex,label={right:$O$}] at (leaf30) {};
\node[vertex,label={below left:$D$}] at (leaf31) {};
\node[vertex,label={below:$E$}] at (leaf32) {};
\node[vertex,label={below right:$O$}] at (leaf33) {};
\end{tikzpicture}%
}
\par\vspace{2pt}$H_{4}$
\end{minipage}\hfill%
\begin{minipage}[t]{.32\linewidth}
\centering\vspace{0pt}
\resizebox{.98\linewidth}{!}
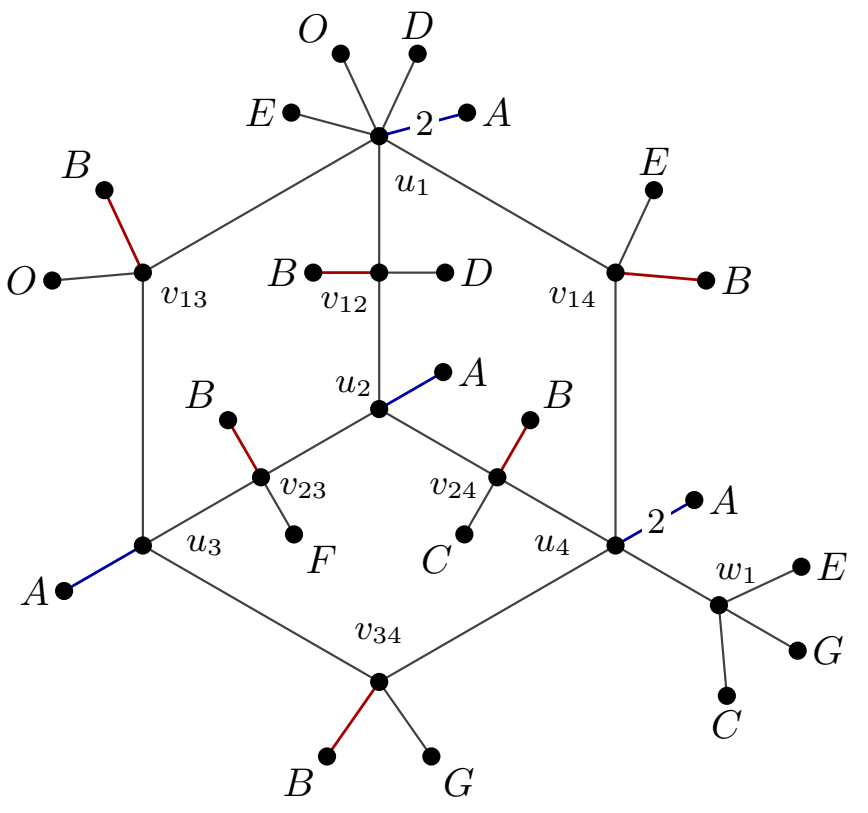
{%
\begin{tikzpicture}[
  x=1cm,y=1cm,
  vertex/.style={circle,fill=black,inner sep=0pt,minimum size=1.6mm},
  edge/.style={draw=black!75,line width=.55pt},
  Aedge/.style={draw=blue!65!black,line width=.7pt},
  Bedge/.style={draw=red!65!black,line width=.7pt},
  weight/.style={fill=white,inner sep=1.2pt,font=\small},
  bulklabel/.style={font=\small,inner sep=0pt},
  every label/.style={font=\normalsize,inner sep=1pt}
]
% H5: polygonal bulk core and repeated boundary leaves.
\coordinate (n8) at (0.00000000,2.40000000);
\coordinate (n13) at (2.07846097,1.20000000);
\coordinate (n9) at (2.07846097,-1.20000000);
\coordinate (n17) at (0.00000000,-2.40000000);
\coordinate (n11) at (-2.07846097,-1.20000000);
\coordinate (n15) at (-2.07846097,1.20000000);
\coordinate (n10) at (0.00000000,0.00000000);
\coordinate (n14) at (0.00000000,1.20000000);
\coordinate (n16) at (1.03923048,-0.60000000);
\coordinate (n18) at (-1.03923048,-0.60000000);
\coordinate (n12) at (2.98778764,-1.72500000);
\coordinate (leaf26) at (3.05751224,-2.52195576);
\coordinate (leaf27) at (3.68060797,-2.12500000);
\coordinate (leaf28) at (3.71283387,-1.38690539);
\coordinate (leaf0) at (0.77274066,2.60705524);
\coordinate (leaf23) at (0.33809461,3.12504623);
\coordinate (leaf24) at (-0.33809461,3.12504623);
\coordinate (leaf25) at (-0.77274066,2.60705524);
\coordinate (leaf2) at (0.56291651,0.32500000);
\coordinate (leaf3) at (-2.77128129,-1.60000000);
\coordinate (leaf1) at (2.77128129,-0.80000000);
\coordinate (leaf5) at (-0.58000000,1.20000000);
\coordinate (leaf30) at (0.58000000,1.20000000);
\coordinate (leaf6) at (-2.41655558,1.92504623);
\coordinate (leaf31) at (-2.87541673,1.13027541);
\coordinate (leaf4) at (2.87541673,1.13027541);
\coordinate (leaf29) at (2.41655558,1.92504623);
\coordinate (leaf9) at (-1.32923048,-0.09770527);
\coordinate (leaf34) at (-0.74923048,-1.10229473);
\coordinate (leaf7) at (1.32923048,-0.09770527);
\coordinate (leaf32) at (0.74923048,-1.10229473);
\coordinate (leaf8) at (-0.45886115,-3.05532164);
\coordinate (leaf33) at (0.45886115,-3.05532164);
\draw[edge] (n8)--(n13);
\draw[edge] (n8)--(n14);
\draw[edge] (n8)--(n15);
\draw[edge] (n9)--(n12);
\draw[edge] (n9)--(n13);
\draw[edge] (n9)--(n16);
\draw[edge] (n9)--(n17);
\draw[edge] (n10)--(n14);
\draw[edge] (n10)--(n16);
\draw[edge] (n10)--(n18);
\draw[edge] (n11)--(n15);
\draw[edge] (n11)--(n17);
\draw[edge] (n11)--(n18);
\draw[Aedge] (n8)--node[weight,pos=.52] {$ 2 $} (leaf0);
\draw[Aedge] (n9)--node[weight,pos=.52] {$ 2 $} (leaf1);
\draw[Aedge] (n10)--(leaf2);
\draw[Aedge] (n11)--(leaf3);
\draw[Bedge] (n13)--(leaf4);
\draw[Bedge] (n14)--(leaf5);
\draw[Bedge] (n15)--(leaf6);
\draw[Bedge] (n16)--(leaf7);
\draw[Bedge] (n17)--(leaf8);
\draw[Bedge] (n18)--(leaf9);
\draw[edge] (n8)--(leaf23);
\draw[edge] (n8)--(leaf24);
\draw[edge] (n8)--(leaf25);
\draw[edge] (n12)--(leaf26);
\draw[edge] (n12)--(leaf27);
\draw[edge] (n12)--(leaf28);
\draw[edge] (n13)--(leaf29);
\draw[edge] (n14)--(leaf30);
\draw[edge] (n15)--(leaf31);
\draw[edge] (n16)--(leaf32);
\draw[edge] (n17)--(leaf33);
\draw[edge] (n18)--(leaf34);
\node[vertex] at (n12) {};
\node[bulklabel] at (3.14778764,-1.44787187) {$ w_{1} $};
\node[vertex] at (n8) {};
\node[bulklabel] at (0.30000000,1.97000000) {$ u_{1} $};
\node[vertex] at (n10) {};
\node[bulklabel] at (-0.22000000,0.20000000) {$ u_{2} $};
\node[vertex] at (n11) {};
\node[bulklabel] at (-1.53000000,-1.20000000) {$ u_{3} $};
\node[vertex] at (n9) {};
\node[bulklabel] at (1.53000000,-1.20000000) {$ u_{4} $};
\node[vertex] at (n14) {};
\node[bulklabel] at (-0.30000000,0.94000000) {$ v_{12} $};
\node[vertex] at (n15) {};
\node[bulklabel] at (-1.70607005,0.98500000) {$ v_{13} $};
\node[vertex] at (n13) {};
\node[bulklabel] at (1.70607005,0.98500000) {$ v_{14} $};
\node[vertex] at (n18) {};
\node[bulklabel] at (-0.66000000,-0.70000000) {$ v_{23} $};
\node[vertex] at (n16) {};
\node[bulklabel] at (0.66000000,-0.70000000) {$ v_{24} $};
\node[vertex] at (n17) {};
\node[bulklabel] at (0.00000000,-1.97000000) {$ v_{34} $};
\node[vertex,label={right:$A$}] at (leaf0) {};
\node[vertex,label={right:$A$}] at (leaf1) {};
\node[vertex,label={right:$A$}] at (leaf2) {};
\node[vertex,label={left:$A$}] at (leaf3) {};
\node[vertex,label={right:$B$}] at (leaf4) {};
\node[vertex,label={left:$B$}] at (leaf5) {};
\node[vertex,label={above left:$B$}] at (leaf6) {};
\node[vertex,label={above right:$B$}] at (leaf7) {};
\node[vertex,label={below left:$B$}] at (leaf8) {};
\node[vertex,label={above left:$B$}] at (leaf9) {};
\node[vertex,label={above:$D$}] at (leaf23) {};
\node[vertex,label={above left:$O$}] at (leaf24) {};
\node[vertex,label={left:$E$}] at (leaf25) {};
\node[vertex,label={below:$C$}] at (leaf26) {};
\node[vertex,label={right:$G$}] at (leaf27) {};
\node[vertex,label={right:$E$}] at (leaf28) {};
\node[vertex,label={above:$E$}] at (leaf29) {};
\node[vertex,label={right:$D$}] at (leaf30) {};
\node[vertex,label={left:$O$}] at (leaf31) {};
\node[vertex,label={below left:$C$}] at (leaf32) {};
\node[vertex,label={below right:$G$}] at (leaf33) {};
\node[vertex,label={below right:$F$}] at (leaf34) {};
\end{tikzpicture}%
}
\par\vspace{2pt}$H_{5}$
\end{minipage}\hfill%
\begin{minipage}[t]{.32\linewidth}
\centering\vspace{0pt}
\resizebox{.98\linewidth}{!}
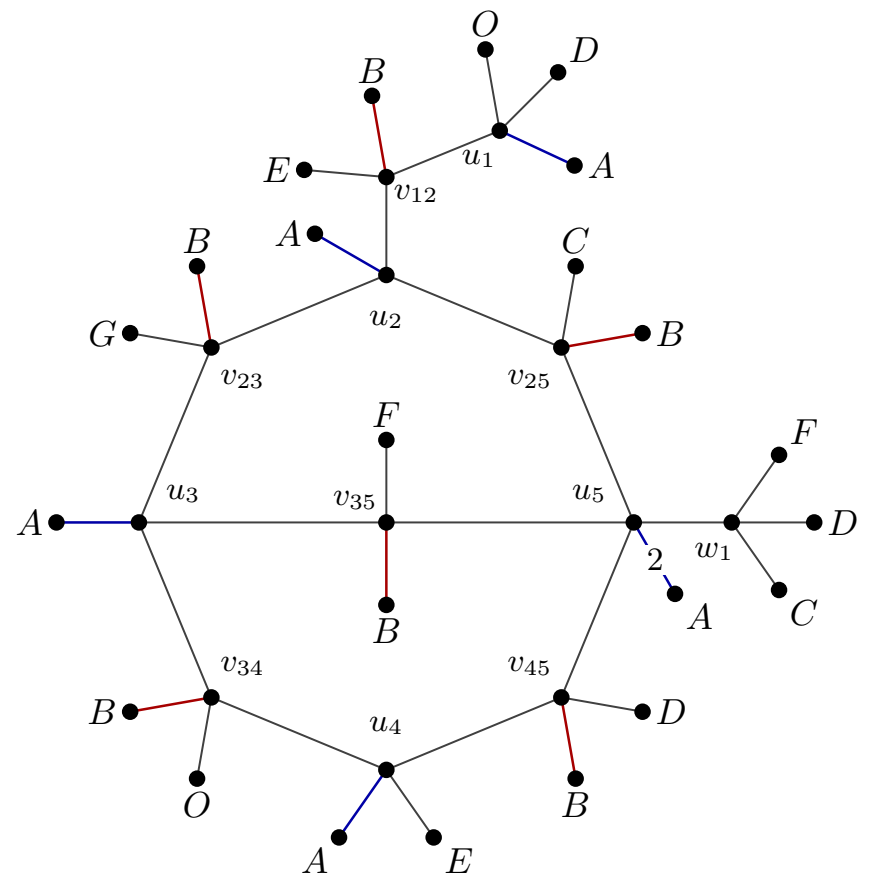
{%
\begin{tikzpicture}[
  x=1cm,y=1cm,
  vertex/.style={circle,fill=black,inner sep=0pt,minimum size=1.6mm},
  edge/.style={draw=black!75,line width=.55pt},
  Aedge/.style={draw=blue!65!black,line width=.7pt},
  Bedge/.style={draw=red!65!black,line width=.7pt},
  weight/.style={fill=white,inner sep=1.2pt,font=\small},
  bulklabel/.style={font=\small,inner sep=0pt},
  every label/.style={font=\normalsize,inner sep=1pt}
]
% H6: polygonal bulk core and repeated boundary leaves.
\coordinate (n9) at (-2.40000000,0.00000000);
\coordinate (n18) at (-1.69705627,1.69705627);
\coordinate (n11) at (0.00000000,2.40000000);
\coordinate (n19) at (1.69705627,1.69705627);
\coordinate (n10) at (2.40000000,0.00000000);
\coordinate (n15) at (1.69705627,-1.69705627);
\coordinate (n12) at (0.00000000,-2.40000000);
\coordinate (n14) at (-1.69705627,-1.69705627);
\coordinate (n17) at (0.00000000,0.00000000);
\coordinate (n13) at (3.35000000,0.00000000);
\coordinate (n16) at (0.00000000,3.35000000);
\coordinate (n8) at (1.10000000,3.80000000);
\coordinate (leaf27) at (3.80886115,-0.65532164);
\coordinate (leaf28) at (4.15000000,0.00000000);
\coordinate (leaf29) at (3.80886115,0.65532164);
\coordinate (leaf0) at (1.82504623,3.46190539);
\coordinate (leaf24) at (1.66568542,4.36568542);
\coordinate (leaf25) at (0.96108146,4.58784620);
\coordinate (leaf3) at (-0.69282032,2.80000000);
\coordinate (leaf1) at (-3.20000000,0.00000000);
\coordinate (leaf4) at (-0.45886115,-3.05532164);
\coordinate (leaf26) at (0.45886115,-3.05532164);
\coordinate (leaf2) at (2.80000000,-0.69282032);
\coordinate (leaf7) at (-0.13891854,4.13784620);
\coordinate (leaf32) at (-0.79695576,3.41972459);
\coordinate (leaf9) at (-1.83597482,2.48490248);
\coordinate (leaf34) at (-2.48490248,1.83597482);
\coordinate (leaf10) at (2.48490248,1.83597482);
\coordinate (leaf35) at (1.83597482,2.48490248);
\coordinate (leaf5) at (-2.48490248,-1.83597482);
\coordinate (leaf30) at (-1.83597482,-2.48490248);
\coordinate (leaf8) at (-0.00000000,-0.80000000);
\coordinate (leaf33) at (0.00000000,0.80000000);
\coordinate (leaf6) at (1.83597482,-2.48490248);
\coordinate (leaf31) at (2.48490248,-1.83597482);
\draw[edge] (n8)--(n16);
\draw[edge] (n9)--(n14);
\draw[edge] (n9)--(n17);
\draw[edge] (n9)--(n18);
\draw[edge] (n10)--(n13);
\draw[edge] (n10)--(n15);
\draw[edge] (n10)--(n17);
\draw[edge] (n10)--(n19);
\draw[edge] (n11)--(n16);
\draw[edge] (n11)--(n18);
\draw[edge] (n11)--(n19);
\draw[edge] (n12)--(n14);
\draw[edge] (n12)--(n15);
\draw[Aedge] (n8)--(leaf0);
\draw[Aedge] (n9)--(leaf1);
\draw[Aedge] (n10)--node[weight,pos=.52] {$ 2 $} (leaf2);
\draw[Aedge] (n11)--(leaf3);
\draw[Aedge] (n12)--(leaf4);
\draw[Bedge] (n14)--(leaf5);
\draw[Bedge] (n15)--(leaf6);
\draw[Bedge] (n16)--(leaf7);
\draw[Bedge] (n17)--(leaf8);
\draw[Bedge] (n18)--(leaf9);
\draw[Bedge] (n19)--(leaf10);
\draw[edge] (n8)--(leaf24);
\draw[edge] (n8)--(leaf25);
\draw[edge] (n12)--(leaf26);
\draw[edge] (n13)--(leaf27);
\draw[edge] (n13)--(leaf28);
\draw[edge] (n13)--(leaf29);
\draw[edge] (n14)--(leaf30);
\draw[edge] (n15)--(leaf31);
\draw[edge] (n16)--(leaf32);
\draw[edge] (n17)--(leaf33);
\draw[edge] (n18)--(leaf34);
\draw[edge] (n19)--(leaf35);
\node[vertex] at (n13) {};
\node[bulklabel] at (3.18000000,-0.28000000) {$ w_{1} $};
\node[vertex] at (n8) {};
\node[bulklabel] at (0.90000000,3.53000000) {$ u_{1} $};
\node[vertex] at (n11) {};
\node[bulklabel] at (0.00000000,1.97000000) {$ u_{2} $};
\node[vertex] at (n9) {};
\node[bulklabel] at (-1.97000000,0.28000000) {$ u_{3} $};
\node[vertex] at (n12) {};
\node[bulklabel] at (0.00000000,-1.97000000) {$ u_{4} $};
\node[vertex] at (n10) {};
\node[bulklabel] at (1.97000000,0.28000000) {$ u_{5} $};
\node[vertex] at (n16) {};
\node[bulklabel] at (0.29000000,3.19000000) {$ v_{12} $};
\node[vertex] at (n18) {};
\node[bulklabel] at (-1.39300036,1.39300036) {$ v_{23} $};
\node[vertex] at (n19) {};
\node[bulklabel] at (1.39300036,1.39300036) {$ v_{25} $};
\node[vertex] at (n14) {};
\node[bulklabel] at (-1.39300036,-1.39300036) {$ v_{34} $};
\node[vertex] at (n17) {};
\node[bulklabel] at (-0.30000000,0.22000000) {$ v_{35} $};
\node[vertex] at (n15) {};
\node[bulklabel] at (1.39300036,-1.39300036) {$ v_{45} $};
\node[vertex,label={right:$A$}] at (leaf0) {};
\node[vertex,label={left:$A$}] at (leaf1) {};
\node[vertex,label={below right:$A$}] at (leaf2) {};
\node[vertex,label={left:$A$}] at (leaf3) {};
\node[vertex,label={below left:$A$}] at (leaf4) {};
\node[vertex,label={left:$B$}] at (leaf5) {};
\node[vertex,label={below:$B$}] at (leaf6) {};
\node[vertex,label={above:$B$}] at (leaf7) {};
\node[vertex,label={below:$B$}] at (leaf8) {};
\node[vertex,label={above:$B$}] at (leaf9) {};
\node[vertex,label={right:$B$}] at (leaf10) {};
\node[vertex,label={above right:$D$}] at (leaf24) {};
\node[vertex,label={above:$O$}] at (leaf25) {};
\node[vertex,label={below right:$E$}] at (leaf26) {};
\node[vertex,label={below right:$C$}] at (leaf27) {};
\node[vertex,label={right:$D$}] at (leaf28) {};
\node[vertex,label={above right:$F$}] at (leaf29) {};
\node[vertex,label={below:$O$}] at (leaf30) {};
\node[vertex,label={right:$D$}] at (leaf31) {};
\node[vertex,label={left:$E$}] at (leaf32) {};
\node[vertex,label={above:$F$}] at (leaf33) {};
\node[vertex,label={left:$G$}] at (leaf34) {};
\node[vertex,label={above:$C$}] at (leaf35) {};
\end{tikzpicture}%
}
\par\vspace{2pt}$H_{6}$
\end{minipage}\par\vspace{7pt}
\end{minipage}
\end{center}

\subsection{Witness inequalities}
\label{app:hec7-witnesses}
\noindent
Without necessarily confirming the extremality of the graphs in Appendix~\ref{app:hec7-graphs}, it is particularly useful to find non-superbalanced holographic inequalities that are saturated by their entropy vectors. This produces pairs $(G,e)$ such as Figure~\ref{fig:witness} and inequality~(\ref{eq:witnessineq}) in the main text, which satisfy \eqref{eq:criterion-bell}, \eqref{eq:criterion-saturation}, \eqref{eq:criterion-mutualinfo}. Each such inequality, decomposed into facets, contains at least one non-superbalanced facet of the holographic entropy cone. 

The relevant inequalities, indexed in the way that corresponds to entropy vectors $s(H_r)$ ($1 \leq r \leq 6$), are given below. Each of them has $\pair{b_{AB}}{e}=2$. We have sought inequalities with few terms on the greater-than side of the inequality, so as to ensure that the contraction maps which validate them remain concise. 

\begin{equation}
\begin{aligned}
 &4S_{ABCE}+7S_{ACEF}+6S_{BDEF}+6S_{BCEG}+5S_{ADEG}\\
 &\geq S_{A}+5S_{B}+3S_{ACE}+S_{ACDE}+3S_{BCDE}+2S_{BCDEF}+S_{BCG}\\
 &+S_{DG}+S_{ADG}+3S_{ABCEG}+S_{ACFG}+S_{BCFG}+2S_{ADFG}+S_{BDFG}.
\end{aligned}
\label{app:hec7-hei-1}
\end{equation}
\begin{equation}
\begin{aligned}
 &8S_{ABDE}+12S_{ACEF}+10S_{BDEF}+8S_{BCEG}+10S_{ADEG}\\
 &\geq 8S_{B}+6S_{ADE}+2S_{BCDE}+2S_{ACF}+2S_{AEF}+2S_{BEF}+4S_{ABDEF}\\
 &+4S_{BCDEG}+6S_{ACFG}+2S_{BCFG}+2S_{BDFG}+2S_{AEFG}.
\end{aligned}
\label{app:hec7-hei-2}
\end{equation}
\begin{equation}
\begin{aligned}
 &5S_{ACDE}+4S_{ABDF}+6S_{BCDF}+4S_{BCEG}+5S_{ACFG}\\
 &\geq 4S_{B}+S_{BCD}+2S_{ACE}+S_{BCF}+2S_{ADF}+4S_{ABCDF}+S_{ADEF}\\
 &+2S_{BDEF}+2S_{ACG}+S_{ADFG}+2S_{BDFG}.
\end{aligned}
\label{app:hec7-hei-3}
\end{equation}
\begin{equation}
\begin{aligned}
 &5S_{ACDE}+4S_{BCDF}+4S_{BCEG}+5S_{ACFG}\\
 &\geq 4S_{B}+2S_{AC}+S_{ACDF}+S_{BCEF}+S_{ADEF}+2S_{AEG}+S_{BDEG}\\
 &+2S_{AEFG}+2S_{BCEFG}.
\end{aligned}
\label{app:hec7-hei-4}
\end{equation}
\begin{equation}
\begin{aligned}
 &7S_{ACDE}+7S_{BCDF}+9S_{BCEG}+8S_{ACFG}+6S_{CEFG}\\
 &\geq 7S_{B}+S_{CE}+2S_{ACE}+3S_{ACEF}+S_{BCEF}+S_{ADEF}+2S_{BDEF}\\
 &+3S_{AEG}+4S_{CFG}+S_{ACDFG}+3S_{AEFG}+S_{BEFG}+5S_{BCEFG}.
\end{aligned}
\label{app:hec7-hei-5}
\end{equation}
\begin{equation}
\begin{aligned}
 &7S_{ACDE}+9S_{BCDF}+7S_{BCEG}+8S_{ACFG}+6S_{CDFG}\\
 &\geq 7S_{B}+S_{CD}+2S_{ACD}+3S_{ADF}+3S_{ACDG}+S_{BCDG}+S_{ADEG}\\
 &+2S_{BDEG}+4S_{CFG}+3S_{ADFG}+S_{BDFG}+5S_{BCDFG}+S_{ACEFG}.
\end{aligned}
\label{app:hec7-hei-6}
\end{equation}
\medskip
% For $r=1,\ldots,6$, let $e_r$ denote the corresponding inequality in
% Eqs.~\eqref{app:hec7-hei-1}--\eqref{app:hec7-hei-6}.
% The normalization in Eq.~\eqref{app:hec7-hei-2} is twice the primitive
% integer normalization of that HEI, so all six pairs satisfy the
% main-text criterion with:
% \begin{equation*}
%  \langle H_r,e_r\rangle=0,\qquad
%  \langle b_{AB},e_r\rangle=2,\qquad I_{H_r}(A:B)=2.
% \end{equation*}

\noindent
Contraction maps, which verify the validity of these inequalities are shown below. 
The input bits correspond to distinct LHS regions in their displayed order.
The output bits follow the RHS regions in their displayed order, with coefficients reflected by multiple bits.
\medskip

% Compact contraction-map tables at a consistent digit size.
\begingroup
% REVTeX normally switches longtable to \small; keep these digits at 10pt.
\expandafter\def\csname table@hook\endcsname{\fontsize{10}{9.6}\selectfont}
\fontsize{10}{9.6}\selectfont
\renewcommand{\arraystretch}{1}
\setlength{\tabcolsep}{0pt}
\setlength{\LTpre}{4pt}
\setlength{\LTpost}{4pt}
\setlength{\LTleft}{\fill}
\setlength{\LTright}{\fill}

% HEC7 contraction maps f_1 and f_2
\begin{longtable}{@{}r@{\hspace{6pt}}l@{\hspace{24pt}}r@{\hspace{6pt}}l@{}}
\toprule
\multicolumn{2}{c}{\textbf{Map $f_1$}} & \multicolumn{2}{c}{\textbf{Map $f_2$}}\\
\cmidrule(r{12pt}){1-2}\cmidrule(l{12pt}){3-4}
$x$ & $f_1(x)$ & $x$ & $f_2(x)$\\
\midrule
\endfirsthead
\toprule
\multicolumn{2}{c}{\textbf{Map $f_1$}\,\textnormal{(continued)}} & \multicolumn{2}{c}{\textbf{Map $f_2$}\,\textnormal{(continued)}}\\
\cmidrule(r{12pt}){1-2}\cmidrule(l{12pt}){3-4}
$x$ & $f_1(x)$ & $x$ & $f_2(x)$\\
\midrule
\endhead
\bottomrule
\endfoot
\bottomrule
\endlastfoot
\texttt{00000} & \texttt{00000000000000000000000000} & \texttt{00000} & \texttt{000000000000000000000000000000000000000000}\\
\texttt{00001} & \texttt{00000000000000001100000111} & \texttt{00001} & \texttt{000000000000000000000010001110111000001011}\\
\texttt{00010} & \texttt{00000000000000010011101001} & \texttt{00010} & \texttt{000000000000001000000000001111000000111000}\\
\texttt{00011} & \texttt{00000000000000011111111111} & \texttt{00011} & \texttt{000000000000000000000000001111111111111111}\\
\texttt{00100} & \texttt{00000000001111100000000001} & \texttt{00100} & \texttt{000000000000000000001111110000000000111100}\\
\texttt{00101} & \texttt{00000000011111101100000111} & \texttt{00101} & \texttt{000000000000001000000011111110000000001100}\\
\texttt{00110} & \texttt{01000000001111110011101001} & \texttt{00110} & \texttt{100000000000001100001111111111000000111000}\\
\texttt{00111} & \texttt{00000000001111111111101111} & \texttt{00111} & \texttt{000000000000000000000011101111100000101110}\\
\texttt{01000} & \texttt{00000000010001000011011100} & \texttt{01000} & \texttt{000000000000000011110000000000111111110000}\\
\texttt{01001} & \texttt{10000000010000000111010110} & \texttt{01001} & \texttt{000000000000000011110011101000111111000011}\\
\texttt{01010} & \texttt{00000000011101110011111000} & \texttt{01010} & \texttt{000000000000001111000000001111111111110000}\\
\texttt{01011} & \texttt{00000000010001100011111110} & \texttt{01011} & \texttt{000000000000001010110011101111111111100011}\\
\texttt{01100} & \texttt{00000000000001100000011111} & \texttt{01100} & \texttt{000000000000000011111111110000111111111111}\\
\texttt{01101} & \texttt{00000000011001100011010111} & \texttt{01101} & \texttt{000000000000001000111111111110111110001011}\\
\texttt{01110} & \texttt{00000000001111100011011101} & \texttt{01110} & \texttt{000000000000001111111111111111111111111010}\\
\texttt{01111} & \texttt{00000000011111100011100100} & \texttt{01111} & \texttt{000000000000001100111111111111110000000011}\\
\texttt{10000} & \texttt{00000000000000000011101000} & \texttt{10000} & \texttt{000000000000000000000011111100000000001010}\\
\texttt{10001} & \texttt{00000000000000000111100111} & \texttt{10001} & \texttt{000000001111100000000011111110111000001011}\\
\texttt{10010} & \texttt{00000000001111010011101001} & \texttt{10010} & \texttt{100000000000001000000011111111000000111100}\\
\texttt{10011} & \texttt{00000000000001010011111101} & \texttt{10011} & \texttt{000000001000000000000011111111111100101111}\\
\texttt{10100} & \texttt{00000000001111100011101001} & \texttt{10100} & \texttt{000000000000001100001111111111000000111100}\\
\texttt{10101} & \texttt{00000000011111100111100111} & \texttt{10101} & \texttt{000000001111111100000011111111000000001100}\\
\texttt{10110} & \texttt{01111100001111110011101001} & \texttt{10110} & \texttt{111111110000001100001111111111000000111100}\\
\texttt{10111} & \texttt{00000000001111110011101001} & \texttt{10111} & \texttt{000000001000001100001111111111000000101100}\\
\texttt{11000} & \texttt{00000011110001000011111100} & \texttt{11000} & \texttt{000000000000000010110011111000111111100010}\\
\texttt{11001} & \texttt{10000011110000000111110110} & \texttt{11001} & \texttt{000000001111110011110011110000111111000011}\\
\texttt{11010} & \texttt{00000011111111110011111000} & \texttt{11010} & \texttt{000000000000001110110011111111111110110000}\\
\texttt{11011} & \texttt{00000011110001100011111100} & \texttt{11011} & \texttt{000000001111101010110011111111111110000011}\\
\texttt{11100} & \texttt{00000010000001100011111111} & \texttt{11100} & \texttt{000000000000001000111111111110111110111011}\\
\texttt{11101} & \texttt{00000011111001100011110111} & \texttt{11101} & \texttt{000000001111111000111011111111111110001011}\\
\texttt{11110} & \texttt{00000010001111100011101001} & \texttt{11110} & \texttt{000000000000001100111111111111000000111010}\\
\texttt{11111} & \texttt{00000011111111100011100000} & \texttt{11111} & \texttt{000000001111111100111111111111000000000011}\\
\end{longtable}

% HEC7 contraction map f_3, in 2 parallel input ranges.
\par\smallskip
{\centering
\noindent
\begin{tabular}{@{}r@{\hspace{6pt}}l@{\hspace{24pt}}r@{\hspace{6pt}}l@{}}
\toprule
\multicolumn{4}{c}{\textbf{Map $f_3$}}\\
\cmidrule(r{6pt}){1-2}\cmidrule(l{6pt}){3-4}
$x$ & $f_3(x)$ & $x$ & $f_3(x)$\\
\midrule
\texttt{00000} & \texttt{0000000000000000000000} & \texttt{10000} & \texttt{0000010000110011000000}\\
\texttt{00001} & \texttt{0000000000110000010110} & \texttt{10001} & \texttt{0000011000110010011100}\\
\texttt{00010} & \texttt{0000000000000001100011} & \texttt{10010} & \texttt{0000011000000011100000}\\
\texttt{00011} & \texttt{0000000000000000011111} & \texttt{10011} & \texttt{0000011000000000011000}\\
\texttt{00100} & \texttt{0000000000111101000010} & \texttt{10100} & \texttt{0000110000111111100110}\\
\texttt{00101} & \texttt{0000000100111111010111} & \texttt{10101} & \texttt{0000110100111110010100}\\
\texttt{00110} & \texttt{0000100100111101100011} & \texttt{10110} & \texttt{0000111000111101100000}\\
\texttt{00111} & \texttt{0000000100111100011011} & \texttt{10111} & \texttt{0000111100111100011000}\\
\texttt{01000} & \texttt{0000000000111100000000} & \texttt{11000} & \texttt{0000010011111111000000}\\
\texttt{01001} & \texttt{0000000011111100010110} & \texttt{11001} & \texttt{0000011011111110011100}\\
\texttt{01010} & \texttt{0000000000111101100011} & \texttt{11010} & \texttt{0000011000111111100000}\\
\texttt{01011} & \texttt{0000000000111100011111} & \texttt{11011} & \texttt{0000011000111100011000}\\
\texttt{01100} & \texttt{0000100100111101100011} & \texttt{11100} & \texttt{0000100011111111100111}\\
\texttt{01101} & \texttt{0000000111111111100111} & \texttt{11101} & \texttt{0000100111111111010100}\\
\texttt{01110} & \texttt{1111100100111101100011} & \texttt{11110} & \texttt{0000100000111101100011}\\
\texttt{01111} & \texttt{0000000100111101100011} & \texttt{11111} & \texttt{0000100100111101010000}\\
\bottomrule
\end{tabular}\par
}

% HEC7 contraction map f_4, in 4 parallel input ranges.
\par\smallskip
{\centering
\noindent
\begin{tabular}{@{}r@{\hspace{6pt}}l@{\hspace{12pt}}r@{\hspace{6pt}}l@{\hspace{12pt}}r@{\hspace{6pt}}l@{\hspace{12pt}}r@{\hspace{6pt}}l@{}}
\toprule
\multicolumn{8}{c}{\textbf{Map $f_4$}}\\
\cmidrule(r{6pt}){1-2}\cmidrule(l{6pt}r{6pt}){3-4}\cmidrule(l{6pt}r{6pt}){5-6}\cmidrule(l{6pt}){7-8}
$x$ & $f_4(x)$ & $x$ & $f_4(x)$ & $x$ & $f_4(x)$ & $x$ & $f_4(x)$\\
\midrule
\texttt{0000} & \texttt{0000000000000000} & \texttt{0100} & \texttt{0000000100010011} & \texttt{1000} & \texttt{0000000011111000} & \texttt{1100} & \texttt{0000001010010000}\\
\texttt{0001} & \texttt{0000001001001110} & \texttt{0101} & \texttt{0000001110001111} & \texttt{1001} & \texttt{0000111011101100} & \texttt{1101} & \texttt{0000111010001010}\\
\texttt{0010} & \texttt{0000000100010011} & \texttt{0110} & \texttt{1111000100010011} & \texttt{1010} & \texttt{0000000111111111} & \texttt{1110} & \texttt{0000000110010011}\\
\texttt{0011} & \texttt{0000000001111111} & \texttt{0111} & \texttt{0000000100011011} & \texttt{1011} & \texttt{0000111101101111} & \texttt{1111} & \texttt{0000111100000011}\\
\bottomrule
\end{tabular}\par
}

% HEC7 contraction maps f_5 and f_6
\par\smallskip
\begin{longtable}{@{}r@{\hspace{6pt}}l@{\hspace{24pt}}r@{\hspace{6pt}}l@{}}
\toprule
\multicolumn{2}{c}{\textbf{Map $f_5$}} & \multicolumn{2}{c}{\textbf{Map $f_6$}}\\
\cmidrule(r{12pt}){1-2}\cmidrule(l{12pt}){3-4}
$x$ & $f_5(x)$ & $x$ & $f_6(x)$\\
\midrule
\endfirsthead
\toprule
\multicolumn{2}{c}{\textbf{Map $f_5$}\,\textnormal{(continued)}} & \multicolumn{2}{c}{\textbf{Map $f_6$}\,\textnormal{(continued)}}\\
\cmidrule(r{12pt}){1-2}\cmidrule(l{12pt}){3-4}
$x$ & $f_5(x)$ & $x$ & $f_6(x)$\\
\midrule
\endhead
\bottomrule
\endfoot
\bottomrule
\endlastfoot
\texttt{00000} & \texttt{0000000000000000000000000000000000} & \texttt{00000} & \texttt{0000000000000000000000000000000000}\\
\texttt{00001} & \texttt{0000000000000000000000000110011110} & \texttt{00001} & \texttt{0000000000000000000000001100111100}\\
\texttt{00010} & \texttt{0000000000110010000000001111100000} & \texttt{00010} & \texttt{0000000000000110010000001111000001}\\
\texttt{00011} & \texttt{0000000000110010000011001111111110} & \texttt{00011} & \texttt{0000000000000110010011001111111101}\\
\texttt{00100} & \texttt{0000000000000101100000000000111111} & \texttt{00100} & \texttt{0000000000000000001100000001111100}\\
\texttt{00101} & \texttt{0000000000000101111100000111111111} & \texttt{00101} & \texttt{0000000000000000011110001101111111}\\
\texttt{00110} & \texttt{0000000000000000010000001111111111} & \texttt{00110} & \texttt{0000000000000111011100001111111101}\\
\texttt{00111} & \texttt{0000000000000000011111111111111111} & \texttt{00111} & \texttt{0000000000000111111111111111111111}\\
\texttt{01000} & \texttt{0000000000000001100000000000111110} & \texttt{01000} & \texttt{0000000000000000101100000001111110}\\
\texttt{01001} & \texttt{0000000000000011100010001110111111} & \texttt{01001} & \texttt{0000000000111000101100001111111110}\\
\texttt{01010} & \texttt{0000000000111011100000001111111110} & \texttt{01010} & \texttt{0000000000100000000000001111111111}\\
\texttt{01011} & \texttt{0000000000111111100011111111111111} & \texttt{01011} & \texttt{0000000000111000000011111111111111}\\
\texttt{01100} & \texttt{1111111000000101100000000000111111} & \texttt{01100} & \texttt{1111111000000000101100000001111110}\\
\texttt{01101} & \texttt{1000000000000101100000000000111111} & \texttt{01101} & \texttt{1000000000000000101100000001111110}\\
\texttt{01110} & \texttt{0000000000000101100000001000111111} & \texttt{01110} & \texttt{0000000000000000101100000001111111}\\
\texttt{01111} & \texttt{0000000000000100000011111000111111} & \texttt{01111} & \texttt{0000000000000000100011110001111111}\\
\texttt{10000} & \texttt{0000000000111011000000000110000000} & \texttt{10000} & \texttt{0000000000000111011000001100000000}\\
\texttt{10001} & \texttt{0000000000111011010000000111011110} & \texttt{10001} & \texttt{0000000000100111011000001110111100}\\
\texttt{10010} & \texttt{0000000011111010011100001111000000} & \texttt{10010} & \texttt{0000000011111111010000001110000001}\\
\texttt{10011} & \texttt{0000000011111010011010001111011110} & \texttt{10011} & \texttt{0000000011110111010010001110111101}\\
\texttt{10100} & \texttt{0000000010111111100000000110111111} & \texttt{10100} & \texttt{0000000000000000011100000000000001}\\
\texttt{10101} & \texttt{0000000111111111111100000111111111} & \texttt{10101} & \texttt{0000000000000110011100000000111101}\\
\texttt{10110} & \texttt{0000000011111100011000001111111111} & \texttt{10110} & \texttt{0000000010000111011000001100100001}\\
\texttt{10111} & \texttt{0000000111111100011111111111111111} & \texttt{10111} & \texttt{0000000010000111011011001100111111}\\
\texttt{11000} & \texttt{0000000000000011100000001000000000} & \texttt{11000} & \texttt{0000000010000111111100001101111110}\\
\texttt{11001} & \texttt{0000000000110011100000001000011110} & \texttt{11001} & \texttt{0000000111111111111100001111111110}\\
\texttt{11010} & \texttt{0000000010111011000000001110010000} & \texttt{11010} & \texttt{0000000011110111100000001111111111}\\
\texttt{11011} & \texttt{0000000010111011000011001110011111} & \texttt{11011} & \texttt{0000000111111111100011111111111111}\\
\texttt{11100} & \texttt{0000000000000101100000000000111111} & \texttt{11100} & \texttt{0000000000000000101100000001111110}\\
\texttt{11101} & \texttt{0000000111111101100000000000111111} & \texttt{11101} & \texttt{0000000111000111101100000001111110}\\
\texttt{11110} & \texttt{0000000011111101000000001000011111} & \texttt{11110} & \texttt{0000000011000111101000000000111111}\\
\texttt{11111} & \texttt{0000000111111100000011111000011111} & \texttt{11111} & \texttt{0000000111000111100011110000111111}\\
\end{longtable}
\endgroup

\subsection{Non-superbalanced facets}
\label{app:facet-examples}
% Center a shared-width block so all inequality starts and signs align.
\newlength{\HECfacetwidth}
\setlength{\HECfacetwidth}{470pt}
% Self-contained display box: no mathtools dependency in the main preamble.
\newcommand{\HECfacetbox}[1]{%
  \makebox[\HECfacetwidth][l]{\(\displaystyle #1\)}%
}
% The seven inequalities below give explicit, certified representatives
% of distinct HEC$_7$ NSNS facet orbits under the full $S_8$ permutation
% group. The last is the 42-term facet $F_1\geq0$ discussed in the main text.
% Here $\mathcal B_7=\{A,B,C,D,E,F,G,O\}$, with $O$ the purifier,
% and purity gives $S_X=S_{\mathcal B_7\setminus X}$.

\noindent
For each inequality~(\ref{app:hec7-hei-1}-\ref{app:hec7-hei-6}), we display the non-superbalanced facet in its facet expansion. The seventh inequality below is similarly extracted from inequality~(\ref{eq:witnessineq}) in the main text.
\begingroup
\small
\setlength{\jot}{1pt}

\begin{equation}
\HECfacetbox{%
\begin{aligned}
 &S_{DE}+S_{FG}+S_{AO}+S_{CDO}+3S_{CEF}+S_{CEO}+3S_{CGO}+3S_{DGO}+2S_{ADG}+2S_{AEF}+S_{BCO}\\
 &+4S_{BEF}+5S_{ACDE}+5S_{ACFG}+4S_{BCDF}+4S_{BCEG}\\
 &\geq 5S_{C}+4S_{D}+2S_{E}+3S_{F}+2S_{G}+S_{O}+S_{A}+S_{CE}+2S_{CG}+S_{CO}+2S_{DO}\\
 &+2S_{EF}+S_{EO}+4S_{GO}+S_{AD}+S_{AE}+S_{AF}+S_{AG}+4S_{BE}+4S_{BF}+S_{BO}+2S_{ADE}\\
 &+2S_{ADO}+2S_{AFG}+2S_{AGO}+S_{ACDG}+3S_{ACEF}.
\end{aligned}
}
\label{app:hec7-facet-0}
\end{equation}

\begin{equation}
\HECfacetbox{%
\begin{aligned}
 &S_{AC}+S_{AO}+2S_{CDG}+6S_{CDO}+2S_{CEF}+3S_{DGO}+5S_{EFG}+2S_{EFO}+S_{ADG}+2S_{AEF}+S_{BDG}\\
 &+3S_{BEF}+6S_{ACDE}+6S_{ACFG}+6S_{BCDF}+6S_{BCEG}\\
 &\geq 3S_{C}+3S_{E}+3S_{F}+5S_{G}+2S_{O}+3S_{A}+2S_{B}+5S_{CD}+2S_{CE}+S_{CG}+S_{DG}\\
 &+6S_{DO}+4S_{EF}+S_{EG}+S_{FG}+2S_{FO}+2S_{GO}+S_{AE}+S_{AF}+S_{BD}+3S_{BE}+3S_{BF}\\
 &+S_{BG}+3S_{ACD}+2S_{ACF}+S_{ACG}+3S_{ADO}+2S_{AEO}+S_{AGO}+4S_{AEFG}.
\end{aligned}
}
\label{app:hec7-facet-1}
\end{equation}

\begin{equation}
\HECfacetbox{%
\begin{aligned}
 &S_{AC}+S_{AO}+S_{CDG}+6S_{CDO}+2S_{CEF}+3S_{DGO}+5S_{EFG}+2S_{EFO}+S_{ADG}+2S_{AEF}+2S_{BDG}\\
 &+3S_{BEF}+6S_{ACDE}+6S_{ACFG}+6S_{BCDF}+6S_{BCEG}\\
 &\geq 4S_{C}+3S_{E}+3S_{F}+5S_{G}+2S_{O}+3S_{A}+S_{B}+4S_{CD}+2S_{CE}+S_{DG}+6S_{DO}\\
 &+4S_{EF}+S_{EG}+S_{FG}+2S_{FO}+2S_{GO}+S_{AE}+S_{AF}+2S_{BD}+3S_{BE}+3S_{BF}+2S_{BG}\\
 &+3S_{ACD}+2S_{ACF}+S_{ACG}+3S_{ADO}+2S_{AEO}+S_{AGO}+4S_{AEFG}.
\end{aligned}
}
\label{app:hec7-facet-2}
\end{equation}

\begin{equation}
\HECfacetbox{%
\begin{aligned}
 &S_{CD}+S_{AD}+S_{AG}+S_{CEF}+S_{CFO}+S_{DEF}+S_{DEG}+4S_{DGO}+4S_{EFG}+S_{ACO}+2S_{AEF}\\
 &+5S_{BCO}+5S_{ACDE}+5S_{ACFG}+4S_{BCDF}+4S_{BCEG}\\
 &\geq 2S_{C}+5S_{D}+3S_{E}+4S_{F}+3S_{G}+S_{O}+2S_{A}+S_{CE}+S_{CO}+2S_{DE}+S_{DO}\\
 &+S_{EF}+S_{EG}+3S_{FG}+3S_{GO}+2S_{AC}+S_{AE}+S_{AF}+4S_{BC}+5S_{BO}+S_{ACD}+S_{ACF}\\
 &+2S_{ADO}+2S_{AEG}+S_{ADEF}+2S_{AEFG}+S_{BCEF}+S_{BDEG}.
\end{aligned}
}
\label{app:hec7-facet-3}
\end{equation}

\begin{equation}
\HECfacetbox{%
\begin{aligned}
 &S_{CD}+S_{AD}+S_{AG}+2S_{CEF}+S_{CFO}+2S_{DEF}+S_{DEG}+4S_{DGO}+3S_{EFG}+S_{ACO}+2S_{AEF}\\
 &+5S_{BCO}+5S_{ACDE}+5S_{ACFG}+4S_{BCDF}+4S_{BCEG}\\
 &\geq S_{C}+5S_{D}+2S_{E}+4S_{F}+4S_{G}+S_{O}+2S_{A}+2S_{CE}+S_{CF}+S_{CO}+3S_{DE}\\
 &+S_{DO}+2S_{EF}+2S_{FG}+3S_{GO}+3S_{AC}+S_{AE}+S_{AF}+4S_{BC}+5S_{BO}+2S_{ADO}+2S_{AEG}\\
 &+S_{ACDF}+S_{ADEF}+2S_{AEFG}+S_{BCEF}+S_{BDEG}.
\end{aligned}
}
\label{app:hec7-facet-5}
\end{equation}

\begin{equation}
\HECfacetbox{%
\begin{aligned}
 &3S_{EG}+2S_{CDG}+2S_{CDO}+S_{CEO}+5S_{CFO}+3S_{DGO}+2S_{EFG}+2S_{EFO}+S_{ACO}+3S_{ADG}+2S_{AEF}\\
 &+7S_{BEF}+7S_{ACDE}+7S_{ACFG}+7S_{BCDF}+7S_{BCEG}\\
 &\geq 5S_{C}+6S_{D}+5S_{E}+S_{F}+9S_{G}+5S_{O}+S_{A}+S_{CD}+S_{CE}+2S_{CF}+S_{CG}\\
 &+S_{CO}+S_{DO}+2S_{EF}+S_{EO}+5S_{FO}+S_{GO}+2S_{AD}+S_{AE}+S_{AF}+2S_{AG}+7S_{BE}\\
 &+7S_{BF}+3S_{ACD}+3S_{ACF}+3S_{ADO}+3S_{AEG}+2S_{AEFG}+2S_{BCDG}+S_{BCEF}+S_{BDEF}.
\end{aligned}
}
\label{app:hec7-facet-6}
\end{equation}

\begin{equation}
\HECfacetbox{%
\begin{aligned}
 &5(S_{ACDE}+S_{ACFG}+S_{EFG}+S_{BCO})+4(S_{BCDF}+S_{BCEG}+S_{DGO})
  +S_{AD}+S_{CD}+S_{AEF}\\
 &+S_{CEF}+S_{DEF}+S_{AG}+S_{DEG}+S_{CFO}+S_{ACO}\\
 &\geq 5(S_D+S_{BO})+4(S_{BC}+S_F+S_{FG})+3(S_A+S_E+S_{GO})
  +2(S_C+S_{AC}+S_{DE}+S_G)\\
 &+2(S_{EG}+S_{AEG}+S_{AEFG}+S_{ADO})+S_{ACD}+S_{CE}+S_{ACF}
  +S_{EF}+S_{BCEF}+S_{ADEF}\\
 &+S_{BDEG}+S_{DO}+S_{CO}+S_O.
\end{aligned}
}
\label{app:hec7-facet-f1}
\end{equation}

\endgroup

\end{document}